\documentclass[aps,prb,twocolumn,superscriptaddress,nofootinbib]{revtex4-2}
\usepackage{epsfig}
\usepackage{subfigure}
\usepackage[T1]{fontenc} 
\usepackage{float}
\usepackage{graphicx}
\usepackage{dcolumn}
\usepackage{bm}
\usepackage{color}
\usepackage{amsmath}
\usepackage[color=orange!40]{todonotes}

\usepackage{amssymb}
\usepackage{xcolor}
\usepackage{braket}
\usepackage[colorlinks = true,
            linkcolor = blue,
            urlcolor  = red,
            citecolor = blue,
            anchorcolor = blue]{hyperref}
\usepackage{mathdots}

\begin{document}
\title{Ground state and spectral properties of the doped one-dimensional optical Hubbard-Su-Schrieffer-Heeger model}
\author{Debshikha Banerjee}
\affiliation{Department of Physics and Astronomy, The University of Tennessee, Knoxville, Tennessee 37996, USA}
\affiliation{Institute for Advanced Materials and Manufacturing, University of Tennessee, Knoxville, Tennessee 37996, USA\looseness=-1}
\author{Jinu Thomas}
\affiliation{Department of Physics and Astronomy, The University of Tennessee, Knoxville, Tennessee 37996, USA}
\affiliation{Institute for Advanced Materials and Manufacturing, University of Tennessee, Knoxville, Tennessee 37996, USA\looseness=-1}
\author{Alberto Nocera}
\affiliation{Stewart Blusson Quantum Matter Institute, University of British Columbia,
Vancouver, British Columbia, V6T 1Z4 Canada}
\affiliation{Department of Physics Astronomy, University of British Columbia, Vancouver, British Columbia V6T 1Z1, Canada}
\author{Steven Johnston}
\affiliation{Department of Physics and Astronomy, The University of Tennessee, 
Knoxville, Tennessee 37996, USA}
\affiliation{Institute for Advanced Materials and Manufacturing, University of Tennessee, Knoxville, Tennessee 37996, USA\looseness=-1}

\begin{abstract}
We present a density matrix renormalization group (DMRG) study of the doped one-dimensional (1D) Hubbard-Su-Schrieffer-Hegger (Hubbard-SSH) model, where the atomic displacements linearly modulate the nearest-neighbor hopping integrals. Focusing on an optical variant of the model in the strongly correlated limit relevant for cuprate spin chains, we examine how the SSH interaction modifies the model's ground and excited state properties. The SSH coupling weakly renormalizes the model's single- and two-particle response functions for electron-phonon ($e$-ph) coupling strengths below a parameter-dependent critical value $g_\mathrm{c}$. For larger $e$-ph coupling, the sign of the effective hopping integrals changes for a subset of orbitals, which drives a lattice dimerization distinct from the standard nesting-driven picture in 1D. The spectral weight of the one- and two-particle dynamical response functions are dramatically rearranged across this transition, with significant changes in the ground state correlations. We argue that this dimerization results from the breakdown of the linear approximation for the $e$-ph coupling and thus signals a fundamental limitation of the linear SSH interaction. Our results have consequences for our understanding of how SSH-like interactions can enter the physics of strongly correlated quantum materials, including the recently synthesized doped cuprate spin chains. 
\end{abstract}

\maketitle
\section{Introduction}
In the Su-Schrieffer-Hegger (SSH) model for polyacetylene~\cite{SSH}, the atomic motion modulates the nearest-neighbor hopping integrals, leading to an electron-phonon ($e$-ph) interaction that is off-diagonal in orbital space. This microscopic coupling mechanism thus modulates the electron's kinetic energy as opposed to its potential energy as in the canonical Holstein~\cite{Holstein} or Fr{\"o}hlich~\cite{Frohlich} models. While the SSH model in one-dimension (1D) has been studied extensively since its inception, interest in SSH-like interactions\footnote{The SSH model initially dealt with acoustic phonons. Here, we use the term ``SSH interaction'' to describe the case where the hopping integral depends on the distance between the atoms to linear order, independent of whether the relevant phonon branch is acoustic or optical. This type of coupling mechanism is sometimes referred to as a Peierls coupling in the literature.} in a broader class of models and systems has recently surged~\cite{Capone1997small, AlexandrovPRB2000, WeberPRB2015, YePRB2016, Moller2017, SousPRL2018, Li2020, XingPRL2021, NoceraPRB2021, SousCoulombGas2022, SousBipolaron2022, fengPRB2022, CohenSteadPreprint, ProkofevPreprint, LiPreprint2022,  Jackiw1976solitons, Heeger1988solitons, JiangPRB2021, Assaad2022, Meier2016observation}. 
This activity has been driven by the realization that such interactions can lead to several novel effects, including the stabilization of nontrivial topological states \cite{Subhajyoti, Jackiw1976solitons, Heeger1988solitons, Meier2016observation} and Dirac points~\cite{Moller2017}, novel bipolaronic charge-density wave~\cite{Li2020, JiangPRB2021, CohenSteadPreprint} and bond-wave~\cite{Assaad2022, fengPRB2022} orderings, and the formation of mobile bipolarons~\cite{SousPRL2018}. There is also a recent proposal that a dilute gas of SSH-bipolarons can have an instability towards a high-temperature (high-T$_\mathrm{c}$) superconducting state~\cite{SousBipolaron2022}. 

Many theoretical studies of the SSH interaction have focused on models without electron correlations. However, SSH couplings are also relevant to many strongly correlated materials. Notable examples include (but are not limited to) quasi-1D~\cite{JohnstonNatureCommun2016, ChenScience2021} and 2D cuprates \cite{Varelogiannis1995quasiparticle, AlexandrovPRB2000, 
Lanzara2001, CukReview2005, ShenPRB2007, Gunnarsson_2008, JohnstonPRB2010, JohnstonPRL2012, PengPRB2022, ChaixNatPhys2017}, manganites~\cite{Blamire1998, MannellaNature2005}, the rare-earth nickelates~\cite{JohnstonPRL2014, Merritt2020}, and other oxides~\cite{MoserPRL2015, FatalePRB2016, CohenSteadPreprint, DrizaNM2012, PhysRevB.95.054510, GASTIASORO2020168107}. In many of these examples, the relevant modes involve the bond-stretching motion of the transition metal and oxygen atoms, which is naturally described by the SSH coupling mechanism. 

Given the likely ubiquity of SSH-like interactions in strongly correlated systems, studying the correlated SSH models like the single-band Hubbard-SSH model is essential. Nevertheless, only a few non-perturbative studies of the Hubbard-SSH model have been carried out. For example, while the model has been studied in one- and two-dimensions \cite{Pearson2011quantized, Hayden1986correlation, fengPRB2022, Cai2021, LiPreprint2022}, these studies focused on half-filling, where an interesting interplay between the $e$-ph interaction and antiferromagnetism has been observed. Ref.~\cite{fengPRB2022}, for instance, studied the model with optical \emph{bond} phonons and observed a phase transition from a long-range antiferromagnetic state to a bond order wave state at a critical \emph{e}-ph coupling for any non-negative value of the on-site Coulomb repulsion. They also identified an intriguing crossover from a standard Hubbard antiferromagnet (with small electronic kinetic energy and doublon density) to a weak antiferromagnet (characterized by an increased electronic kinetic energy) with strong quantum fluctuations at a weak \emph{e}-ph strength $g^{*}$. 

Far fewer studies have been carried out for doped Hubbard-SSH models. A recent functional renormalization group study of the doped 2D model found evidence for $s$- and $d$-wave superconductivity and spin-density-wave formation for $\rho = 1-\langle n\rangle = 0.15$ hole-doping in the weak SSH coupling limit~\cite{YangRG2022}. A subsequent study on the ground state properties of $\rho = 0.125$ hole-doped four-leg ladders~\cite{WangLadders2022} found similar superconducting correlations at weak coupling and stripe formation at strong coupling. Another recent study has focused on the 1D model's dilute limit using density matrix renormalization group (DMRG)~\cite{NoceraPRB2021}. The authors found that in the absence of Hubbard on-site repulsion, the ground state of the SSH model could be described as a liquid of bipolarons, which remains stable up to large values of the $e$-ph interaction. (The stability of the liquid should be contrasted with the Holstein or extended-Holstein model, where bipolarons are heavy and prone to ordering or phase separation~\cite{Bonca2001,Hohenadler2019dominant, Bradley2021superconducivity, Nosarzewski2021superconductivity}.) However, Ref.~\cite{NoceraPRB2021}  also found that the system was prone to phase separation if the $e$-ph coupling was too large. In this limit, the carriers are separated into a single region of half-filled sites with a bond-wave ordering surrounded by regions of empty sites. Finally, another study of a multi-orbital Hubbard–SSH model in the strongly correlated high-density limit at half-filling also observed lattice instabilities in the strong coupling limit~\cite{LiPreprint2022}. 

The SSH model's tendency towards phase separation in the strong coupling limit can be traced to the breakdown of the linear approximation for the interaction. Empirical fits to \textit{ab initio} electronic structure calculations have shown that the direct nearest-neighbor hopping in many materials scales as $t(\delta d) = t[1+\delta d/d_0]^{-\eta}$, where $d_0$ is the equilibrium bond distance between the atoms, $\delta d$ is the net change in bond length, and $\eta$ ($\approx 2 - 4$) is a positive constant that depends on the angular momentum of the relevant orbitals~\cite{Harrison}. An important feature of this form is that $t(\delta d)$ cannot change sign for any value of $\delta d < d_0$. The linear SSH interaction is obtained by expanding this functional form to first order such that $t(\delta d) \approx t(1 - \tfrac{\eta}{a}\delta d)$. 
Crucially, the effective hopping in this approximation can change sign whenever $\tfrac{\eta}{a}\langle \delta d\rangle > 1$. Since $\eta \approx 2-4$, this condition will be met for $\delta d$ values that are a smaller fraction of the lattice spacing. Therefore, one should regard any sign change in the effective hopping integral as unphysical for most models,\footnote{This discussion assumes that we are concerned with the modulation of the direct hopping between nearest neighbor orbitals. There are situations where an intermediate atom can modulate indirect hopping between orbitals (see, for example, Ref.~\cite{SousBipolaron2022}). In these cases, one can envision cases where the sign change in the effective hopping may be allowed.} and a signal that one should include any additional nonlinear terms in the interaction Hamiltonian.

 Motivated by these considerations, we present a detailed study of the 1D-doped single-band Hubbard-SSH model using DMRG. Our goals are two-fold: first, we would like to study the model's ground state and dynamical correlation functions to understand the effects of these interactions in correlated systems like the cuprates better. Second, we want to identify and understand the consequences of inducing a sign change in the effective hopping in the strong coupling limit, which has yet to be addressed systematically in the literature. To help us realize our first goal, we focus exclusively on the strongly correlated limit ($U = 8t$), and a carrier concentration of $\langle n \rangle = 0.75$ (or $\rho = 0.25$ hole doping). These model parameters are relevant for the recently synthesized doped 1D corner-shared cuprates~\cite{ChenScience2021} and other strongly correlated materials doped away from the Mott insulating regime. To facilitate our second goal, we vary the strength of the $e$-ph coupling $g$ from weak to strong coupling. 
For coupling strengths below a parameter-dependent critical coupling $g_\mathrm{c}$, we find that the SSH interaction weakly dresses the model's static and dynamical correlations. However, for $g > g_\mathrm{c}$, the system develops large displacements that result in the expected sign changes in the effective hopping integral for a subset of the orbitals. This effect subsequently drives a lattice dimerization distinct from the standard weak-coupling Peierl's mechanism. Symptoms of the dimerization also manifest in a dramatic rearrangement of spectral weight in the one- and two-particle dynamical response functions. These results demonstrate that the sign inversion in the effective hopping integral can radically alter the ground state of the model. Since this behavior is also observed in the uncorrelated doped SSH model~\cite{NoceraPRB2021}, our results confirm that it is rooted in the linear approximation for the interaction and should not be underestimated when considering this microscopic coupling mechanism.

\section{Model and methods}
\subsection{The Hubbard-SSH model}
The SSH model was initially introduced to describe acoustic phonons in polyacetylene \cite{SSH}, where the lattice directly modulates the nearest neighbor hopping integrals. A variant of the model, which we call the ``optical'' SSH model, treats the atomic motion using optical phonons while retaining the modulation of the hopping integrals introduced initially in the SSH Hamiltonian~\cite{Capone1997small}.\footnote{
The equivalence of the acoustic SSH model and an optical SSH model in one dimension at half-filling has been established by Weber \textit{et al}. in Ref.~\cite{WeberPRB2015}. However, the model considered by Weber \textit{et al}. defines the phonons to live on the bonds between sites rather than on the sites as in our model. In the bond model, the hopping integrals $t_{i,i+1}$ and $t_{i,i-1}$ are modulated independently. In contrast, in our optical model, \textit{both} hopping integrals are modulated by the motion of the $i$\textsuperscript{th} atom. This distinction is important because no equivalence between the three models has been established, especially away from half-filling.} This model variant is quite appealing for describing materials like transition metal oxides, where the optical bond-stretching motion of the oxygen atoms directly modulates the transition-metal-oxygen hopping integral. 

Here we focus on the optical Hubbard-SSH model in 1D. The model's Hamiltonian is 
\begin{align}\nonumber
    H&=-t \sum_{i,\sigma} \left[
    c^\dagger_{i,\sigma}c^{\phantom\dagger}_{i+1,\sigma} + h.c.\right] + U \sum_{i}\hat{n}_{i,\uparrow}\hat{n}_{i,\downarrow} \\
    &+ \Omega \sum_{i}b^\dagger_i b^{\phantom\dagger}_i 
    + 
   g\sum_{i,\sigma} \left[c^\dagger_{i,\sigma}c^{\phantom\dagger}_{i+1,\sigma}\left(\hat{X}_i - \hat{X}_{i+1}\right) + \text{h.c.}\right],\label{eq:Hamiltonian}
\end{align}
where $c^\dagger_{i,\sigma}$ ($c^{\phantom\dagger}_{i,\sigma}$) creates (annihilates) a spin-$\sigma$ $(=\uparrow,\downarrow)$ electron on lattice site $i$, $b^\dagger_i$ ($b^{\phantom\dagger}_{i,}$) creates (annihilates) a phonon mode with energy $\Omega$ at site $i$, and $\hat{X}_i = (2M\Omega)^{-1/2}(b^\dagger_i + b^{\phantom\dagger}_i)$ is the displacement operator for the atom at site $i$. The remaining model parameters are nearest-neighbor hopping $t$, the on-site Hubbard interaction strength $U$, and the electron-phonon coupling strength $g$. 

Throughout this paper, we consider doped 1D chains of length $L$ with a carrier concentration of $n = 0.75$ $e$/site (or a hole doping $\rho = 1-n = 0.25$). We set $t = 1$ as our unit of energy. To study the effects of correlations that are relevant for materials like the cuprate spin chains, we fix $U = 8t$. We also perform calculations for $\Omega=t$ and $2t$ to vary the degree of retardation in the model.  

\subsection{DMRG and Observables}  

\begin{figure*}[t]
    \centering
    \includegraphics[width=6.5in]{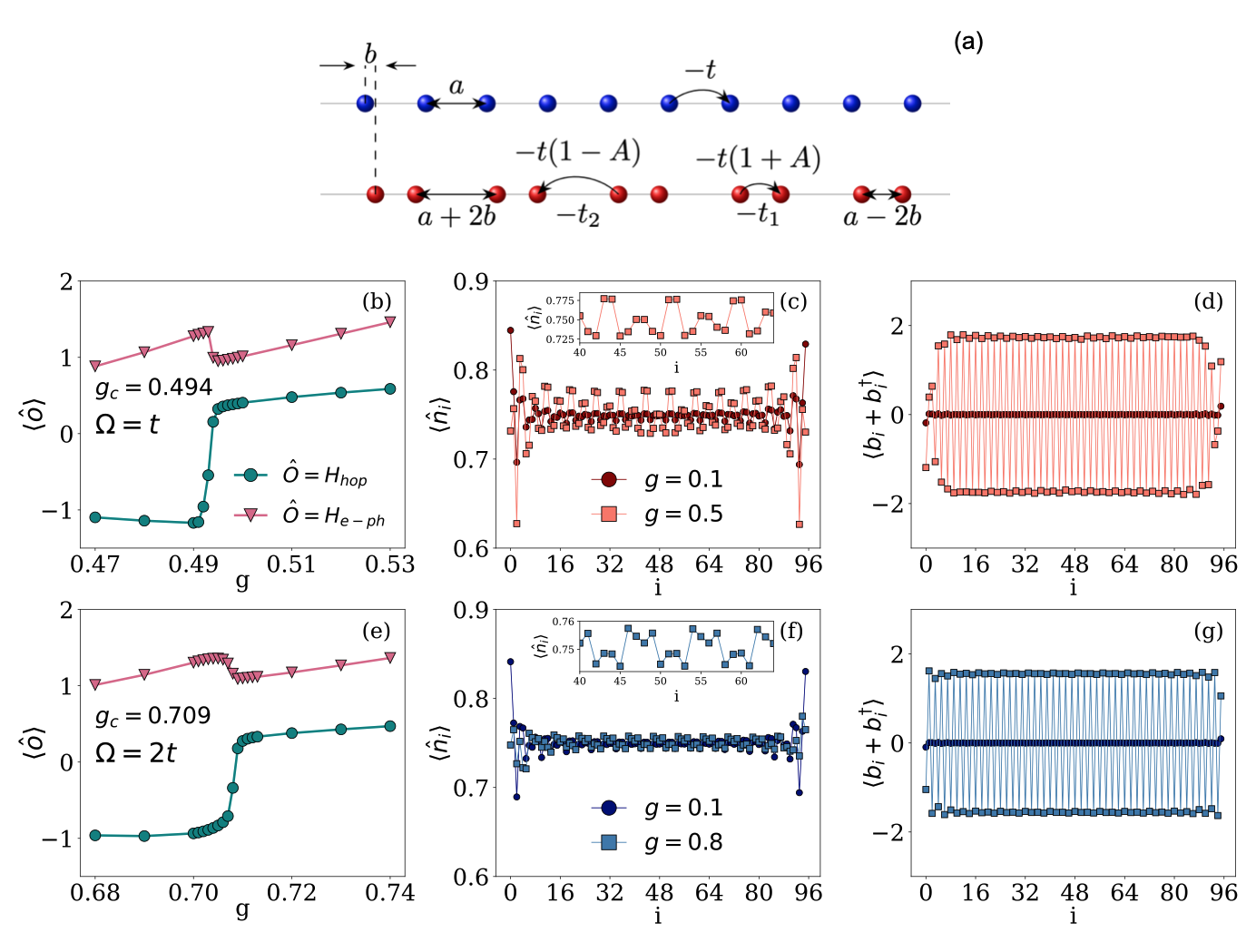}
    \caption{(a) A schematic representation of the two limiting cases of the Hubbard-SSH model treated at the mean-field level. The model reduces to a single-band Hubbard chain with lattice spacing $a$ when the $e$-ph coupling $g = 0$, as shown in the top diagram. When $g>g_\mathrm{c}$, the system enters a dimerized state where the hopping integrals alternate between $-t_1 = -t(1+A)$ and $-t_2 = -t(1-A)$ along the chain direction. 
    (b) A comparison of ground state expectation values of the single-particle hoping $\braket{H_\text{hop}}$ and $e$-ph interaction $\braket{H_{e-\text{ph}}}$ with varying $g$. The expectation is calculated at the center site $c=11$ of a $L=24$ site chain for $U=8t$, $\Omega=t$.
    (c) Ground state expectation values of the average occupation number $\braket{\hat{n}_i}$ with varying $g$. The expectation is calculated at each site of an $L=96$ site chain for $U=8t$, $\Omega=t$. Inset of (c) shows $\braket{\hat{n}_i}$ for $g=0.5>g_\mathrm{c}$ for sites $i=40$ to $64$.
    (d) Ground state expectation values of the lattice distortion $\braket{b_i^{\dagger}+b_i^{\phantom\dagger}}$ with varying $g$. The expectation is calculated at each site of an $L=96$ site chain for $U=8t$, $\Omega=t$.
    Panels (e)-(g) show similar results for $\Omega = 2t$. }
    \label{fig:dimerization}
\end{figure*}

We solved Eq.~\eqref{eq:Hamiltonian} using DMRG \cite{PhysRevLett.69.2863} as implemented in the DMRG++ code~\cite{DMRGpp}. 

To assess the strength of various ordering tendencies, we calculated the ground-state real-space correlation functions $\langle \hat{O}^\dagger_c \hat{O}^{\phantom\dagger}_j\rangle \equiv \bra{\Psi_\mathrm{gs}} \hat{O}^\dagger_c \hat{O}^{\phantom\dagger}_j\ket{\Psi_\mathrm{gs}}$, where $\ket{\Psi_\text{gs}}$ denotes the ground state wavefunction. Here we employ the center-site approximation~\cite{WhitePRL2004}, where $c$ indicates a site in the center of the chain and $j$ is a site on its right half. 

The ground state magnetic and charge correlations are obtained from the spin-spin 
\begin{equation}
C_{\sigma}(r)= \braket{\hat{\bf S}_c \cdot \hat{\bf S}_j}
\end{equation}
and density-density 
\begin{equation}
C_{\rho}(r)=\braket{\hat{n}_{c}\:\hat{n}_{j}}-\braket{\hat{n}_{c}}\braket{\hat{n}_{j}} 
\end{equation}
correlation functions, respectively, where the distance $r=|j-c|$ is measured relative to the central site of the chain. The superconducting correlations are obtained from the pair-pair correlation functions 
\begin{equation}
    C_\text{s(t)} = \braket{\hat{\Delta}_{\text{s(t)},c}^{\dagger}\:\hat{\Delta}^{\phantom\dagger}_{\text{s(t)},j}}, 
\end{equation}
where 
\begin{equation}
\hat{\Delta}_{\text{s},j}^{\dagger}=\frac{1}{\sqrt{2}}[\hat{c}^{\dagger}_{\uparrow,j}\hat{c}^{\dagger}_{\downarrow,j+1}-\hat{c}^{\dagger}_{\downarrow,j}\hat{c}^{\dagger}_{\uparrow,j+1}]
\end{equation}
for spin-singlet pairing and 
\begin{equation}
\hat{\Delta}_{\text{t},j}^{\dagger}=\frac{1}{\sqrt{2}}[\hat{c}^{\dagger}_{\uparrow,j}\hat{c}^{\dagger}_{\downarrow,j+1}+\hat{c}^{\dagger}_{\downarrow,j}\hat{c}^{\dagger}_{\uparrow,j+1}]
\end{equation}
for spin-triplet pairing. 

Unless otherwise stated, our DMRG calculations for ground state correlation functions were carried out on $L=96$ site chains with open boundary conditions. We kept up to $m=1000$ DMRG states to maintain a truncation error below $10^{-7}$ and restricted the local phonon Hilbert space to keep $7$ ($8$) phonon modes per site for phonon energies $\Omega=2t$ ($t$). 
We have checked that our results for the ground state correlation functions are converged with respect to the size of the local phonon Hilbert space. 

We also computed the model's single- and two-particle response functions. Each dynamical correlation function $C_{ij}(\omega)$ is defined in real space using appropriate one- or two-particle operators $\hat{\mathcal{A}}_i$ and $\hat{\mathcal{B}}_i$  
\begin{equation}
    C_{ij}(\omega) =-\frac{1}{\pi}\operatorname{Im}\bra{\Psi_\mathrm{gs}}\hat{\mathcal{A}}_j
    \frac{1}{\omega -\hat{H}+E_{0}+\mathrm{i}\eta}
    \hat{\mathcal{B}}_i^{\phantom\dagger} \ket{\Psi_\mathrm{gs}}. 
    \label{eq:dynamical_corr}
\end{equation}
The corresponding correlation functions in momentum space were then obtained by a Fourier transform. 

The single-particle spectral function $A(k,\omega)$ is calculated from the sum of the electron removal [$A_{ij}^{-}(\omega)$] and addition [$A_{ij}^{+}(\omega)$] spectra, which are defined using the operators 
$\hat{\mathcal{A}}_i = c_{i,\sigma}$, $\hat{\mathcal{B}}^{\phantom\dagger}_j = c^\dagger_{j,\sigma}$ and $\hat{\mathcal{A}}^{\phantom\dagger}_i = c^\dagger_{i,\sigma}$, $\hat{\mathcal{B}}_j = c_{j,\sigma}$, respectively.
The energies of all spectra shown here have been shifted by the chemical potential $\mu=(E_{N+1}-E_{N-1})/2$, where $E_{N}$ is the ground state energy of the system with $N$ particles, to place the Fermi energy at $\omega = 0$. 

We also computed the phonon spectral function $B_{ij}(\omega)$, and the two-particle dynamic spin $S_{ij}(\omega)$ and charge $N_{ij}(\omega)$ structure factors. $B_{ij}(\omega)$ is defined using the operators $\hat{\mathcal{A}}_{i} = \hat{X}_i$ and $\hat{\mathcal{B}}_{i} = (\hat{X}_{i}-\braket{\hat{X}_{c}})$, where $c$ is the center site of the 1D chain. Similarly, the dynamical spin and charge structure factors are defined using the operators $\hat{\mathcal{A}}_{i}=\hat{\mathcal{B}}_{i} = S_{i}^{z}$ and $\hat{\mathcal{A}}_{i}= \hat{n}_{i}$ and $\hat{\mathcal{B}}_{i} = \Tilde{n}_{i}=(\hat{n}_{i}-\braket{\hat{n}_{c}})$, respectively. 

When calculating the dynamical correlation functions, we fixed the broadening coefficient to $\eta = 0.1t$ and computed the spectral functions for each $\omega$ using the Correction-Vector algorithm with Krylov decomposition and a two-site DMRG update~\cite{NoceraPRE2016}, as implemented in the DMRG++ code \cite{DMRGpp}. We kept $m=400$ states and 6 phonon modes per site. To avoid the necessity of re-orthogonalizing the Krylov vectors, we allowed up to 200 Krylov vectors and truncated the effective Hamiltonian decomposition with a tolerance of $10^{-12}$. Our implementation uses a matrix product state representation for the many-body wave function, where the local fermionic and bosonic degrees of freedom are merged into a single index $\sigma$ so that the local physical dimension $d=4\times (n_{\text{ph}}+1)$. (4 is the size of the local fermionic Hilbert space and $n_{\text{ph}}+1$ the size of the local phonon Hilbert space.) 
Although this choice is computationally 
more expensive (see below) and may seem disadvantageous, it allowed us to avoid getting stuck in metastable solutions typical for the 1-site update algorithm. This aspect is important because getting stuck in such metastable solutions appears to occur more frequently in Correction-Vector calculations than ground state calculations~\cite{Hubig2015}. 
On the other end, the most costly operation of the Correction-Vector algorithm is the contraction of the effective Hamiltonian $H^\mathrm{eff}_{\alpha^\prime\beta^\prime\alpha\beta}$ (see Refs.~\cite{NoceraPRB2022,schollwock2011} for more details) with the local 2-site matrix product state tensor $M_{\alpha^\prime\beta^\prime}$. Here we have defined the indices $\alpha=\{m_l,\sigma_i\}$ and $\beta=\{\sigma_{i+1},m_r\}$, with $m_l$ ($m_r$) the left (right) bond dimension index while $\sigma_i$ ($\sigma_{i+1}$) is physical index at site $i$ ($i+1$). This procedure has a computational cost of the order $O(d^3 B^2 m^2 + d^2 B m^3$),  where $B$ is the bond dimension of the Hamiltonian and $m$ the bond dimension of the matrix product state~\cite{schollwock2011}.

\section{Results}

\subsection{Limitations of the linear SSH model}
As discussed in the introduction, the size of the lattice displacements generally increase with the strength of the $e$-ph coupling. Therefore, we expect the average displacements to become large enough to flip the sign of some of the effective hopping integrals $-t_\mathrm{eff} \approx -t+g(X_i-X_{i+1})\equiv  -t+g\delta d_i$ once the coupling is made too large~\cite{NoceraPRB2021, LiPreprint2022}. When this occurs, the system is unstable towards forming a dimerized state~\cite{NoceraPRB2021}, as sketched in Fig.~\ref{fig:dimerization}(a) (Figs.~\ref{fig:dimerization}(d) \& (g) show this phenomenon explicitly, see also discussion below). In this static picture, the hopping along the short bond increases significantly to $-t_1 = -t(1+A)$, where $A \equiv g\delta d > 0$, while the hopping along the long bond decreases to $-t_2 = -t(1-A)$. If $g \delta d_i \gg 1$, then the hopping along the long bond can actually \emph{pass} zero and eventually take large \textit{positive} values, and the magnitude of this positive hopping will continue to grow as $\delta d_i$ increases. This situation is unphysical because the hopping integral $t(\delta d)$ should tend toward zero when the atoms are very far apart. 

Any sign change in $t(\delta d_i)$ along the long bonds should also produce a sudden change in the system's kinetic energy. To confirm this, we calculated 
the ground-state expectation values of the single-particle hopping $\braket{H_\text{hop}} =\langle c^\dagger_{c,\sigma} c^{\phantom\dagger}_{c+1,\sigma}\rangle$ and the $e$-ph interaction $\braket{H_{e-\text{ph}}}=\langle c^\dagger_{c,\sigma} c^{\phantom\dagger}_{c+1,\sigma}(\hat{X}_c-\hat{X}_{c+1}) \rangle$, where $c=11$ is the central site of an $L=24$ site chain. We have found that larger local phonon Hilbert space is generally needed to obtain converged results for $g \approx g_\mathrm{c}$, so here we consider $7$ phonon modes per site for both $\Omega=2t$ and $\Omega=t$. 
The evolution of these quantities as a function of the $e$-ph coupling is plotted in Fig.~\ref{fig:dimerization}(b) \& (e) for $\Omega = t$ and $2t$, respectively. Both undergo a fairly sharp change as the coupling is tuned across a parameter-dependent coupling $g_\mathrm{c}$ ($=0.494$ for $\Omega = t$ and $=0.709$ for $\Omega = 2t$). In particular, the expectation value $\langle H_\text{hop} \rangle$ changes sign at $g_\mathrm{c}$, signaling a large increase in the contribution from the sign-flipped hopping along the long bonds.  

\begin{figure*}[t]
    \begin{minipage}{0.49\textwidth}
    \centering
    \includegraphics[width=\columnwidth]{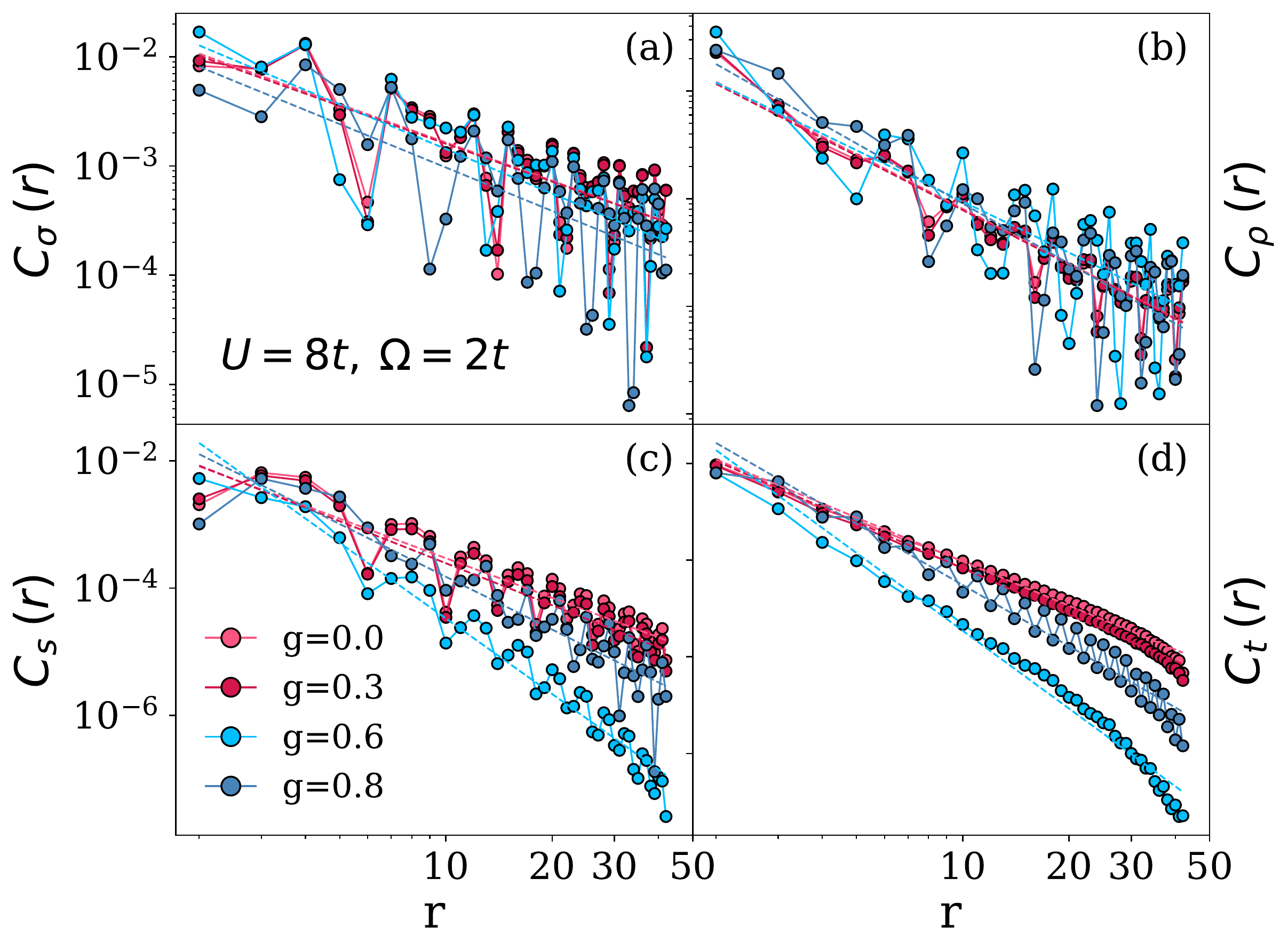}
    \end{minipage}\hfill
    \begin{minipage}{0.49\textwidth}
    \centering
    \includegraphics[width=\columnwidth]{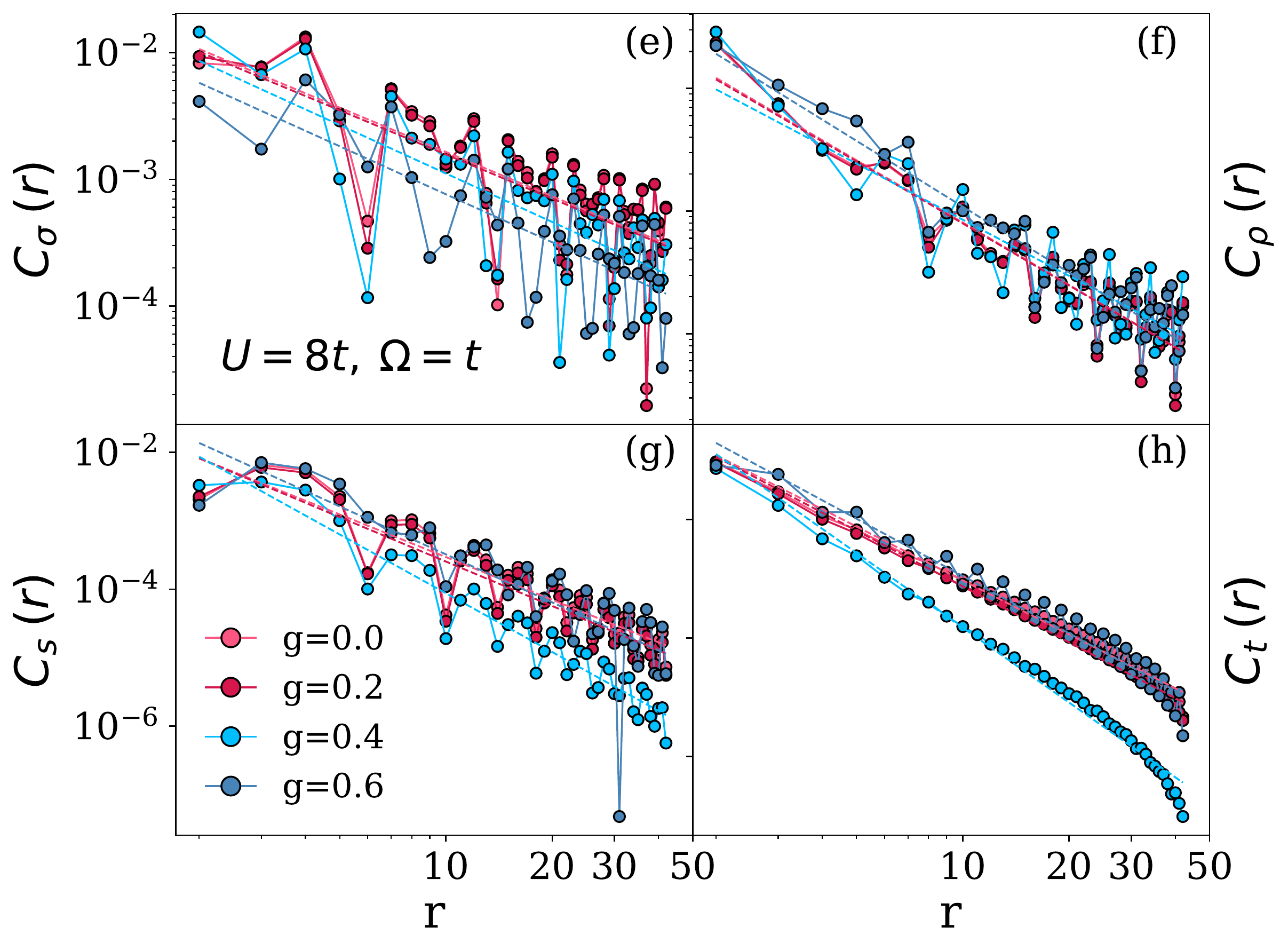}
    \end{minipage}    
   \caption{Log-Log plots of the Hubbard-SSH model's ground state correlation functions plotted as a function of the distance $r=|j-c|$, where $c=48$ is a central site and $j$ is a site along the length of the chain. The individual panels show the (a) spin, (b) charge, and (c) singlet and (d) triplet pairing correlations. These results were obtained for $L=96,\: U=8t,\:\Omega=2t$ and $25\%$ doping. The dashed lines are linear fits of the data, and the legend in panel (c) applies to panels (a)-(d). Here the critical coupling is estimated to be $g_\mathrm{c}\approx0.709$ using an $L=24$ site chain. Panels (e)-(h) show results following the same layout and for the same parameters but with $\Omega = t$. Here the critical coupling is estimated to be $g_\mathrm{c}\approx0.494$ using an $L=24$ site chain. The legend in panel (g) applies to panels (e)-(h). }
    \label{fig:GS_U8_w2}
\end{figure*}

To confirm that these changes in the kinetic energy corresponding to the formation of a dimerized state, we computed the average value of the local electron density $\braket{\hat{n}_i}$ and local lattice displacements $\braket{b_i^\dagger +b^{\phantom\dagger}_i}$, as shown in Figs.~\ref{fig:dimerization}(c), (f) and Figs.~\ref{fig:dimerization}(d),(g), respectively, as a function of position $i$ along the chain direction with varying $g$. Here, panels (c) and (d) show results for $\Omega = t$ while panels (f) and (g) are for $\Omega = 2t$. For a weak $e$-ph coupling of $g=0.1$, $\braket{\hat{n}_i}$  exhibits a weak charge modulation for both values of $\Omega$; however, for $g>g_\mathrm{c}$, the charge density modulations increase such that holes collect on alternating pairs of sites along the chain direction. At the same time, the local displacements  $\braket{b_i^{\dagger}+b^{\phantom\dagger}_i}$ [Fig.\ref{fig:dimerization} (d) \& (g)] transition from an undistorted structure to a dimerized structure for $g>g_\mathrm{c}$, which extends across the entire chain length. Notably, the lattice distortion observed here has a $q = \pi$ wave vector instead of the $q = 2k_\mathrm{F} = 3\pi/4$ structure that is expected based on a weak-coupling nesting-driven Peierls mechanism. This latter observation suggests that the dimerization observed here is a strong-coupling effect. Finally, our results have substantial edge effects on local density and displacements. We have not analyzed these in terms of topological edge states and defer a more detailed study of these effects in future work.

\subsection{Ground state correlations} 

Figure~\ref{fig:GS_U8_w2} shows the spin $C_\sigma(r)$, charge $C_\rho(r)$, and superconducting correlations in the singlet $C_\text{s}(r)$ and triplet $C_\text{t}(r)$ channels as a function of the distance $r=|j-c|$ from the center site of the chain. Results are shown for various couplings $g$ and at a fixed $U = 8t$ and $\Omega=2t$. Similar results for $\Omega = t$ are shown in Fig.~\ref{fig:GS_U8_w2}(e)-(h). 

For the doped Hubbard model, one expects~\cite{giamarchi2004quantum,Capello2005} the correlation functions to decay with $r$ as 
\begin{align}
C_{\sigma}(r)&\sim \frac{K_{\sigma}}{(\pi r)^2}+A^{\sigma}  \frac{\cos(2k_\mathrm{F} r)}{r^{K_{\rho}+1}}\log^{1/2}{(r/\alpha)}\\
C_{\rho}(r)&\sim \frac{K_{\rho}}{(\pi r)^2}+A^{\rho}  \frac{\cos(2k_\mathrm{F} r)}{r^{K_{\rho}+1}} \log^{-3/2}{(r/\alpha)}\\
C_\mathrm{s}(r)&\sim A^\mathrm{s} \frac{\log^{1/2}{(r/\alpha)}}{r^{1/K_{\rho}+1}}\\
C_\mathrm{t}(r)&\sim A^\mathrm{t} \frac{\log^{-3/2}{(r/\alpha)}}{r^{1/K_{\rho}+1}}, 
\end{align}
where we can substitute $r\rightarrow\frac{L}{\pi}\sin(\pi r/L)$ to take into accounts of the effects of the boundary conditions~\cite{Laflorencie2006}. Here, 
$A^{\sigma}, A^{\rho}, A^\mathrm{s}, A^\mathrm{t}$ are non-universal model-dependent constants, and $\alpha$ is a cutoff used to regularize the low energy field theory. In the Hubbard chain away from half-filling $k_\mathrm{F}=\pi n/2$, where $n=N/L$ is electron density and $N=N_{\uparrow}+N_{\downarrow}$. Away from half-filling, it is expected that $K_{\sigma}=1$ while $K_\rho^{-1}=\sqrt{1+\frac{U}{2\pi t}\sin(2k_\mathrm{F})}$; thus, $K_\rho\simeq0.67$ for $U=8t$ $n=0.75$ and $g=0$.
Therefore, the spin and charge correlation functions, modulo log corrections, should decay with the same exponent,\footnote{The same is true for the singlet and triplet superconducting correlations.} and spin/charge power law decay dominate over the superconducting ones. 

\begin{figure}[t]
    \centering
    \includegraphics[width=\columnwidth]{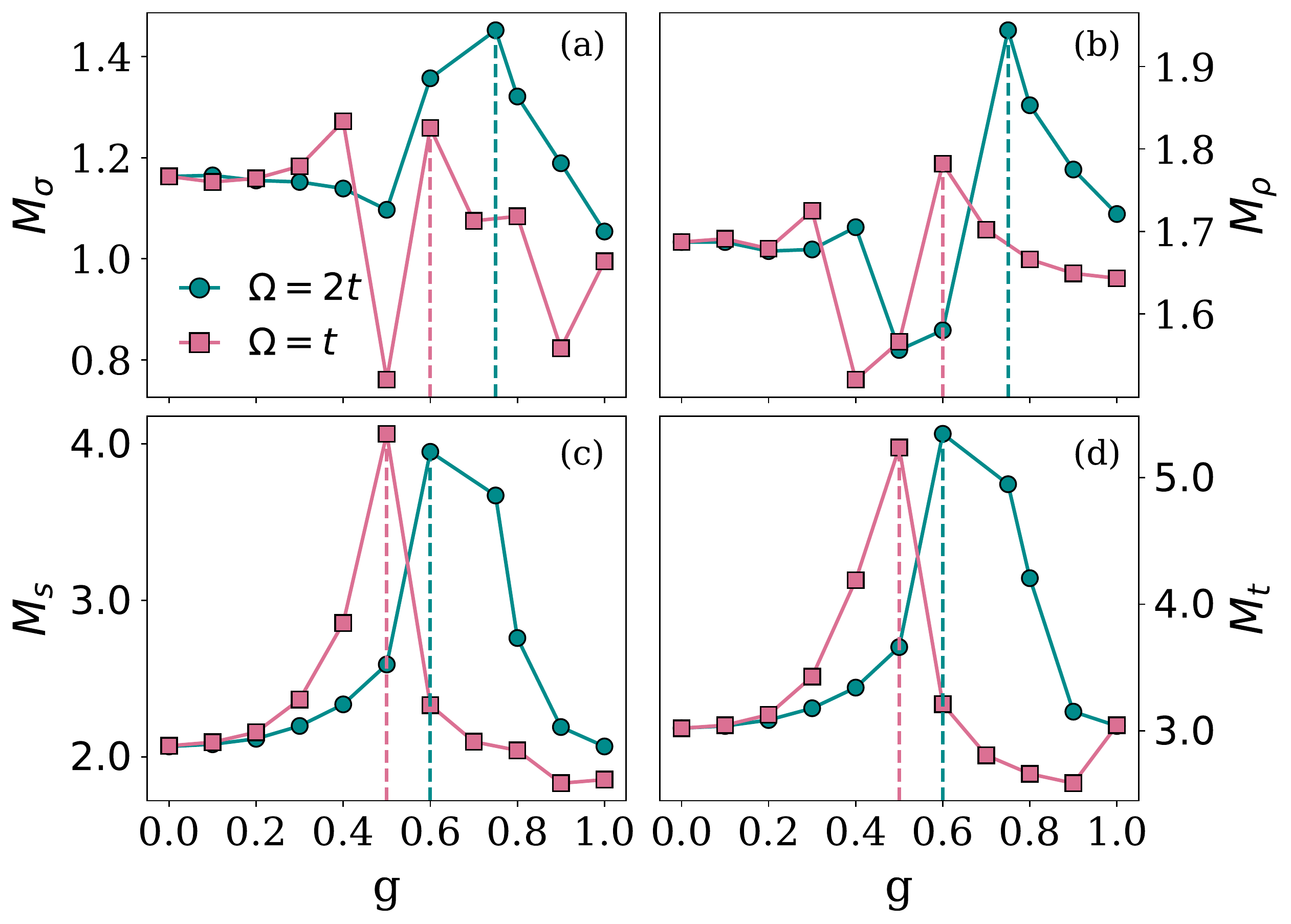}
    \caption{Comparison of decay exponents extracted from various correlation functions for different $e$-ph coupling parameters for $L=96$, $U=8t$, $\Omega=2t$, and $t$. The exponents are extracted by a power law fit $C(r)\sim r^{-M}$ shown in Fig.~\ref{fig:GS_U8_w2}. Here the critical coupling for $L=24$ and $\Omega=2t$ is at $g_\mathrm{c}\approx0.709$ and $g_\mathrm{c}\approx0.494$ for $\Omega=t$. The legend in panel (a) applies to all four panels. }
    \label{fig:slope_U8_w2_w1}
\end{figure}

Extracting the Luttinger liquid parameters for the Hubbard-SSH model is more challenging because the expected form for each correlation function is currently unknown. One might suppose the same form applies in the weak coupling limit, but this will not hold once the system is in the unphysical long-range dimerized state. For this reason, we have not applied any logarithmic corrections to the fitting functions used to analyze the correlation functions and instead plot the absolute value of each $C_\alpha(r)$ as a function of $r$. We then fit the resulting curve with a power law of the form $C_\alpha(r) \sim r^{-M_\alpha}$ ($\alpha = \sigma, \rho, \text{s}, \text{t}$) to determine how the $e$-ph coupling modifies each correlation function. (The dashed lines in each panel indicate the fits.) The evolution of the extracted exponents with $g$ is summarised in Fig.~\ref{fig:slope_U8_w2_w1}. For $g = 0$, we obtain exponents $M_\sigma = 1.163$, $M_\rho = 1.687$, and $M_{\mathrm{s}(\mathrm{t})} = 2.07(3.021)$ for the spin, charge, and singlet (triplet) superconducting correlations. 
For our choice of parameters, field theory predicts $M_{\rho}\approx M_\sigma\approx 1.67$ and $M_\mathrm{s}\approx M_\mathrm{t}\approx 2.49$. Thus, we conclude that the logarithmic corrections and contributions from the uniform part of the spin/charge correlations expected from the field theory are significant. This unfortunate situation prevents us from reliably extracting the Luttinger liquid parameter $K_{\rho}$ for the system size studied here ($L=96$ sites). Nevertheless, the evolution of our extracted exponents carries some information about how the correlations \emph{change} with respect to the case where no SSH \emph{e}-ph is introduced. 
 
The ground state correlations are weakly modified by the $e$-ph coupling for both values of the phonon energy, resulting in slight changes in the exponents $M_\alpha$ for $g \lesssim 0.3$. However, the exponents undergo a more rapid and nonmonotonic change as the coupling increases towards the critical coupling $g_\mathrm{c}$ where dimerization occurs (indicated here by the vertical dashed lines). For example, both the spin [$M_\sigma$, Fig.~\ref{fig:GS_U8_w2}(a)] and charge [$M_\rho$, Fig.~\ref{fig:GS_U8_w2}(b)] exponents initially decrease for just below $g_\mathrm{c}$ before rapidly increasing near $g_\mathrm{c}$ and then ultimately decaying back to smaller values in the strong coupling limit. The corresponding superconducting correlations in both the singlet $C_\mathrm{s}(r)$ [Fig.~\ref{fig:GS_U8_w2}(c)] and triplet correlations $C_\mathrm{t}(r)$ [Fig.~\ref{fig:GS_U8_w2}(d)] channels are also suppressed as $g$ is swept across the critical coupling, resulting in a sharp increase in the exponents $M_\text{s}$ and $M_\text{t}$ by nearly a factor of two before falling back towards their initial values in the strong coupling limit. 

The results in Figs.~\ref{fig:GS_U8_w2}-\ref{fig:slope_U8_w2_w1} demonstrate that the SSH $e$-ph coupling can affect the spin, charge, and superconducting correlations in nontrivial ways. Interestingly, the observed behavior is nonmonotonic, with the most substantial changes in the various correlations occurring near the coupling $g_\mathrm{c}$, where the effective hopping changes sign. This nonmonotonic dependence is a critical difference between the SSH and Holstein or Fr{\"o}hlich models, where sharp but monotonic transitions to the small polaron regime occur in the strong coupling limit~\cite{Bonca1999holstein, Berciu2007systematic, Devreese2009Frohlich, Greitemann2018lecture}. 

Our results show that the most dramatic changes in the ground state correlations can be linked to the lattice fluctuations near the transition to the dimerized state. In this case, the suppression of the superconducting correlations can be associated with the increased tendency to localize pairs on the short bonds and the associated fluctuations of the lattice. 
In the next section, we will examine the dynamical properties of the doped Hubbard-SSH model to understand the spectral signatures of the different regimes identified here.

\subsection{Dynamical correlation functions: limiting cases}
Before turning to the dynamical correlation functions of the Hubbard-SSH model, we first present and discuss the results for two limiting cases to help guide our analysis. The first case is the doped single-band Hubbard model with $L=40$, $U = 8t$, corresponding to our model's $g\rightarrow 0$ limit. The second is a dimerized Hubbard model with $L=40$ and $U = 8t$ but with effective hopping integrals alternating between $-t_1 = -t(1+A)$ and $-t_2 = -t(1-A)$, as shown in Fig.~\ref{fig:dimerization}(a). In this case, we estimate $A$ from a static mean-field-like analysis of the SSH interaction where $A = g\:\langle X_{c}-X_{c+1} \rangle$. Here, $c$ denotes the chains' center site, and the expectation value is evaluated using the ground state of the Hubbard-SSH model obtained with DMRG. We estimate $A$ from the Hubbard-SSH model with $g = 0.6 > g_\mathrm{c}$ and $\Omega = t$, which results in values of $A = 1.46$, $-t_1 = -2.46t$, and $-t_2 = 0.46t$, deep in the dimerized regime. Finally, we fix the carrier concentration to $\langle n \rangle = 0.75$ for both limiting cases. 

\begin{figure}[t]
    \begin{center}
        \includegraphics[width=\columnwidth]{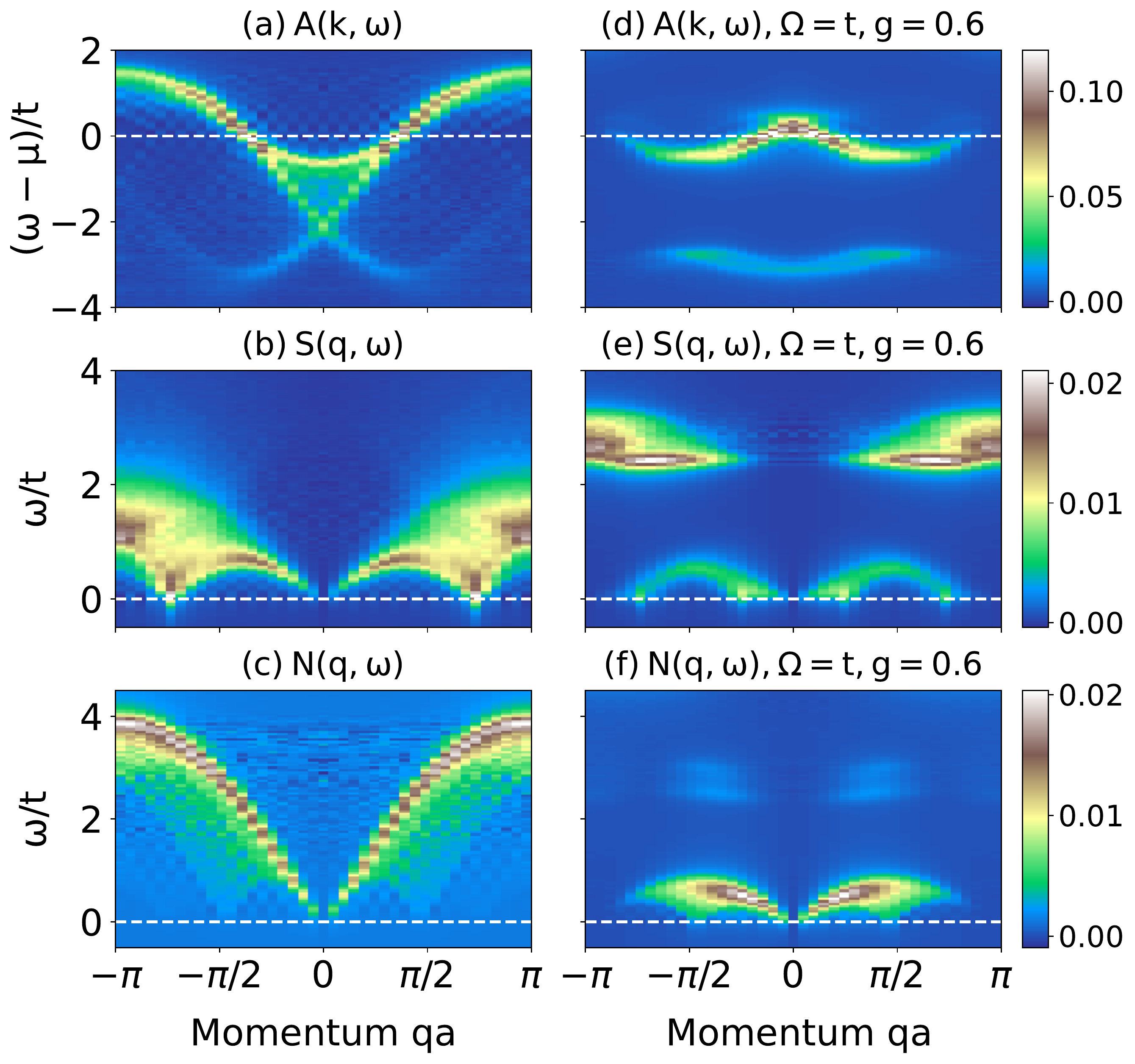}
        \caption{The single-particle spectral function $A(k,\omega)$, spin structure factor $S(q,\omega)$ and charge structure factor $N(q,\omega)$ for the doped Hubbard model with $L=40, U=8t, n=0.75$. Panels (a-c) show results for the Hubbard chain, while panels (d-f) show results for a dimerized model [see Fig.~\ref{fig:dimerization}(a)]. In the latter case, the hopping alternate between $-t_1 = -t(1+A)$ and $-t_2=-t(1-A)$ along the chain as estimated from a mean-field analysis of SSH interaction with $g=0.6$ and $\Omega=t$.}
        \label{fig:limits}
    \end{center}
\end{figure}

Figure~\ref{fig:limits}(a)-(c) show the results for the doped Hubbard model. For this case, the spectral function $A(k,\omega)$ [Fig.~\ref{fig:limits}(a)] agrees well with prior calculations~\cite{PhysRevLett.96.156402, PhysRevB.74.241103, li2021particle}. For example, spin-charge separation is evident in the spectral function from the distinct spinon and holon bands, which form a triangular spectral structure ranging from $-k_\mathrm{F}$ to $k_\mathrm{F}$ ($k_\mathrm{F}=\pi N/2L=0.75 \:\pi/2$). The spin structure factor $S(q,\omega)$ [Fig.~\ref{fig:limits}(b)] also exhibits the typical two spinon continuum for a doped 1D chain~\cite{EderPRB1995, PhysRevB.55.12510, NoceraPRB2016, ParschkePRB2019, li2021particle}. 
For our choice of $U = 8t$, the spectral weight is focused in sharp peaks at the lower boundary of the continuum, corresponding to the Heisenberg limit~\cite{NoceraPRB2016}, while the spectrum is gapless at $q=\pm\: 2k_\mathrm{F}$. Finally, the charge structure factor $N(q,\omega)$ [Fig.~\ref{fig:limits}(c)] exhibits a continuum of excitations with a sharp dispersive feature whose spectral weight concentrated tracks the top of the continuum, again in agreement with prior work~\cite{EderPRB1995, KumarNPJ2018, li2021particle}.

\begin{figure*}[t]
    \begin{minipage}{0.49\textwidth}
    \centering
    \includegraphics[width=\columnwidth]{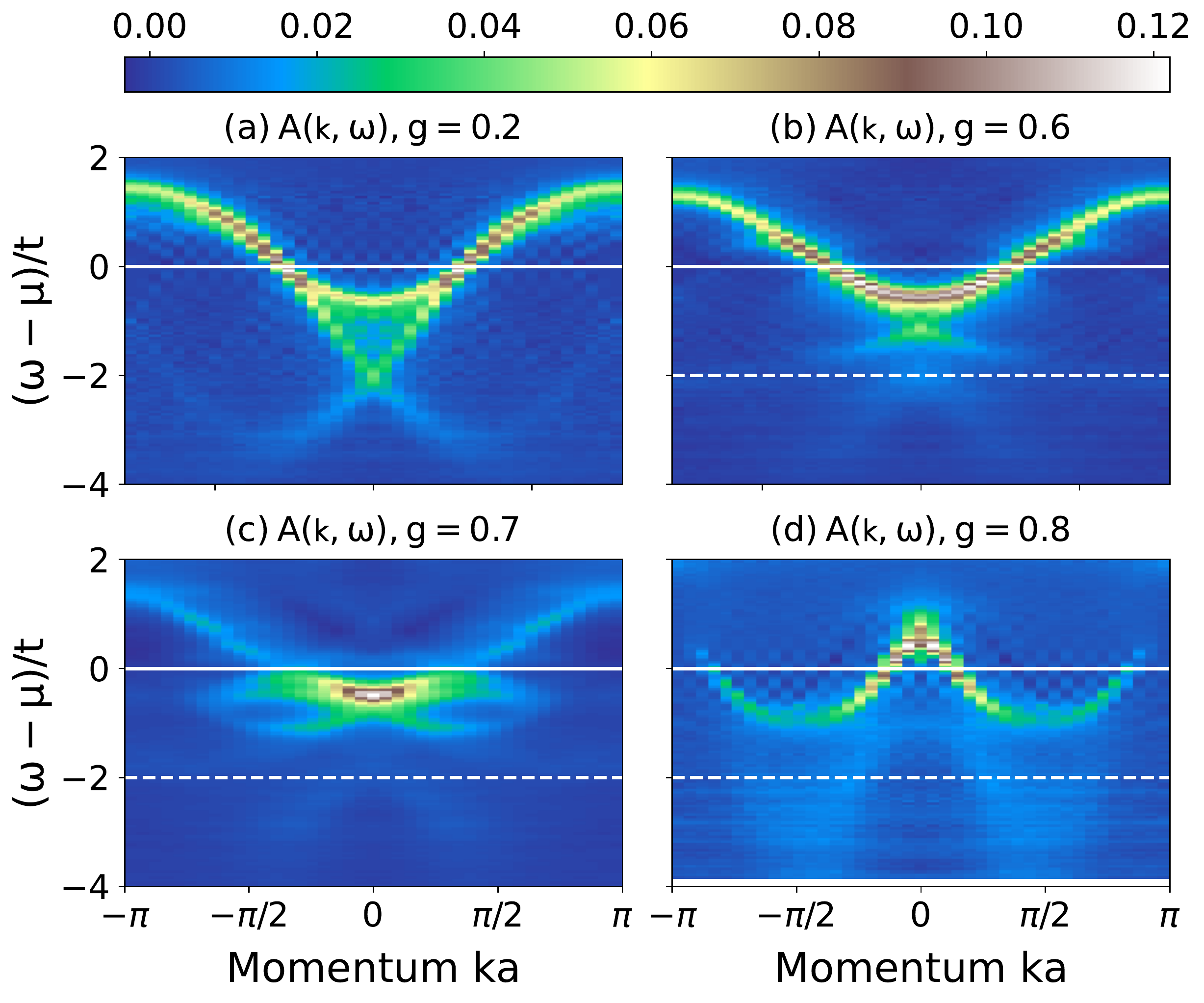}
    \end{minipage}\hfill
    \begin{minipage}{0.49\textwidth}
    \centering
    \includegraphics[width=\columnwidth]{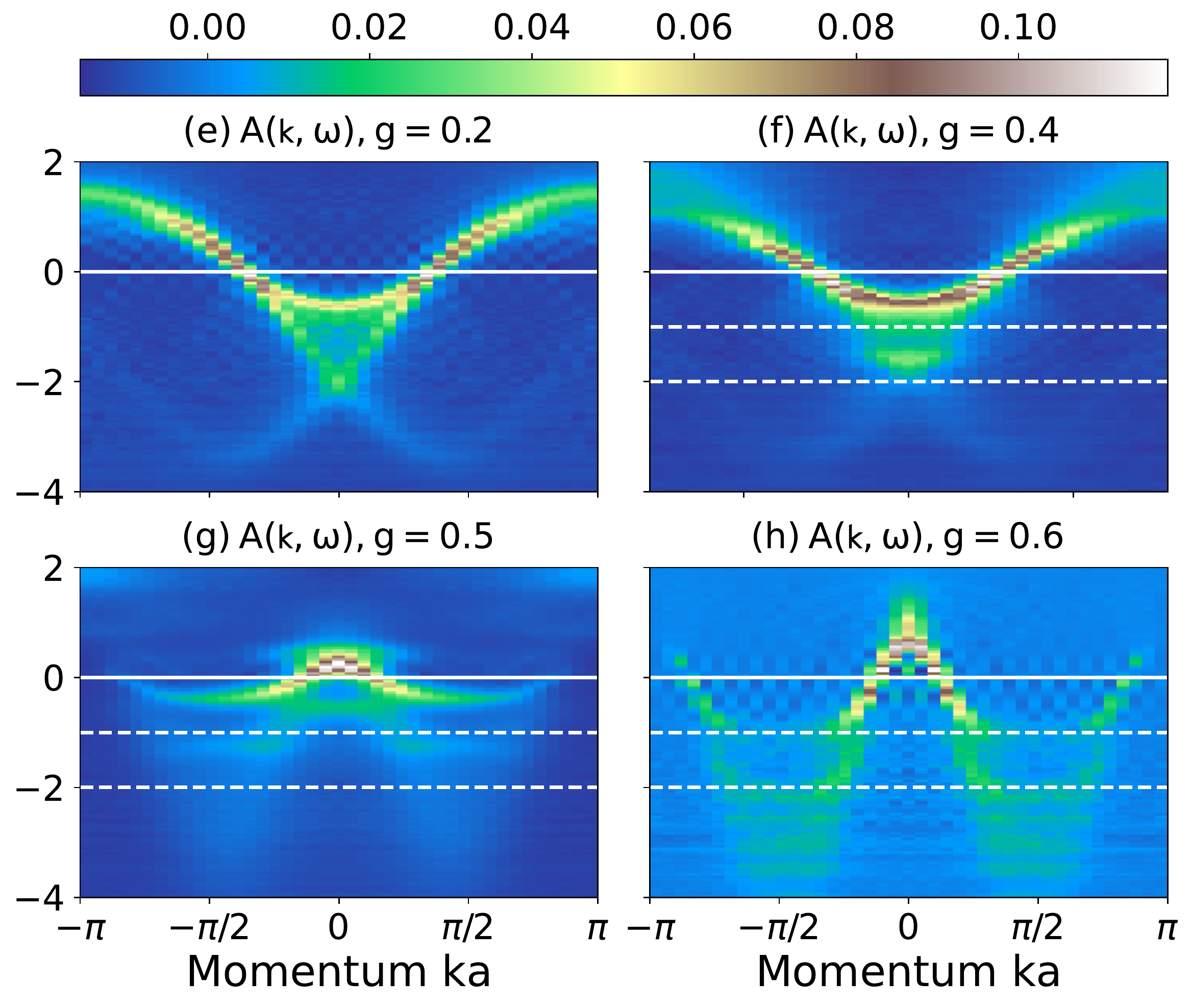}
    \end{minipage}    
    \caption{Single-particle spectral function $A(k,\omega)$ for the doped Hubbard-SSH model. The left set of panels shows results for varying SSH coupling $g$ as indicated in each panel, a fixed $U=8t$ and $\Omega=2t$, and an average filling of $n=0.75$ (corresponding to $25\%$ hole doping). The right set of panels shows similar results for fixed $U=8t$, $\Omega=t$, and $n = 0.75$. All spectra were computed using $L =40$ site chains. For $U=8t, \Omega=2t$ and $L=24$ the critical coupling is at $g_\mathrm{c}\approx0.709$ and for $U=8t, \Omega=t$, $g_\mathrm{c}\approx0.494$.}
    \label{fig:Akw_U8_w2_w1}
\end{figure*}

Figure~\ref{fig:limits}(d)-(f) show $A(k,\omega)$, $S(q,\omega)$, and $N(q,\omega)$ for the dimerized Hubbard model. The spectral weight of the single- and two-particle response functions are completely reorganized in this limit. For example, the main dispersing feature in the spectral function in Fig.~\ref{fig:limits}(a) is now flipped, reflecting the change of sign of the effective hopping $-t_2 = -t(1-A)$. 
As a result, the triangular shape formed from the crossing of spinon and holon bands is inverted and compressed around the Fermi energy. The spectrum also appears to acquire a quasi-periodicity consistent with the enlarged unit cell of the dimerized chain but with the spectral weight differing in the second zone. As a result, the main dispersing feature now has two Fermi surface crossings located at $\pm \: k_\mathrm{F}^{\prime}$ and $\pm(\pi-k_\mathrm{F}^{\prime})$, where $k_\mathrm{F}^{\prime}=3/8\:k_\mathrm{F}$. 

Turning to the two-particle response functions, we find that the spectral weight of the spin and charge excitations are now divided into low- and high-energy branches. For example, the low energy spectral weight of the spin excitations has an inverted parabolic-like shape that crosses $\omega=0$ at several points. In contrast, the high energy weight is relatively dispersionless and concentrated near $q=\pm \pi$. The charge structure factor $N(q,\omega)$ also becomes concentrated at low energies with a sharp dispersive feature that crosses $\omega = 0$ at $q = 0$ and $q\approx \pm\:\pi/2$. A weaker high-energy part also appears, which corresponds to charge fluctuations between the band crossing the Fermi level and the more incoherent states far below $E_\mathrm{F}$ found in Fig.~\ref{fig:limits}(d). In the next section, we will show that the spectral properties of the Hubbard-SSH model interpolate between these two limits as the strength of the $e$-ph coupling increases. 

\subsection{Single particle spectral functions} 

The limiting cases discussed in the previous subsection assume that the lattice distortions are static. In this section, we examine the dynamical properties of the Hubbard-SSH model, where the lattice dynamics are treated fully and on an equal footing as the electron degrees of freedom. 

We begin with the single-particle spectral function $A(k,\omega)$, shown Fig.~\ref{fig:Akw_U8_w2_w1}(a)-(d) for $U=8t$ and $\Omega = 2t$. For weak values of the $e$-ph coupling [$g=0.2$, Figs.~\ref{fig:Akw_U8_w2_w1}(a)], the spectra resemble that of the doped 1D Hubbard model [Fig~\ref{fig:limits}(a)]; the spectra bear the classic signatures of spin-charge separation, and no apparent kinks \cite{SandvikPRB2004, NoceraPRB2014} or other electronic renormalizations can be seen in the data. As $g$ increases, however, various renormalizations become noticeable in the electronic structure at integer multiples of the phonon energy (indicated by the dashed white lines). For example, for $g=0.6$ and $\Omega = 2t$ [Fig~\ref{fig:Akw_U8_w2_w1}(b)], the spectral weight is split up as it crosses the phonon energy and the spectral features above $\Omega$ are pushed upward in energy. This behavior indicates a slight redressing of the carriers by the $e$-ph interaction, leading to an overall increase in the effective mass and a reduction in the holon bandwidth. (This behavior causes the apparent squeezing of the triangular spectral structure.) 
It is important to mention that, for $g<g_\mathrm{c}$, the spectral functions show a spectral feature [see Fig.~\ref{fig:Akw_U8_w2_w1}(b) and (f)] consistent with the so-called holon folding mode at intermediate SSH coupling~\cite{ChenScience2021,WangPRL2021}. However, our data suggests that this feature does not have enough spectral intensity to account for the observed weight in Ba$_{2-x}$Sr$_x$CuO$_{3+\delta}$~\cite{ChenScience2021}.

Once the coupling is increased beyond the critical coupling $g_\mathrm{c}$, we observe a complete reorganization of the spectral weight, consistent with the dimerization of the system. Fig.~\ref{fig:Akw_U8_w2_w1}(d) illustrates this for $g = 0.8 > g_\mathrm{c}$ ($=0.709$ for $U = 8t$ and $\Omega = 2t$). In this case, the spectral function resembles the dimerized limit shown in Fig.~\ref{fig:limits}(d) but with additional incoherent weight at higher binding energies. Fig.~\ref{fig:Akw_U8_w2_w1}(c) shows the spectra for $g=0.7 \lessapprox g_\mathrm{c}$. The spectrum has a mix of features from the dimerized and undimerized cases for this coupling value, suggesting that the system fluctuates between the two states, possibly on short-length and time scales. There are also weak indications of a gap opening near the Fermi level. 

\begin{figure*}[t]
    \begin{minipage}{0.49\textwidth}
    \centering
    \includegraphics[width=\columnwidth]{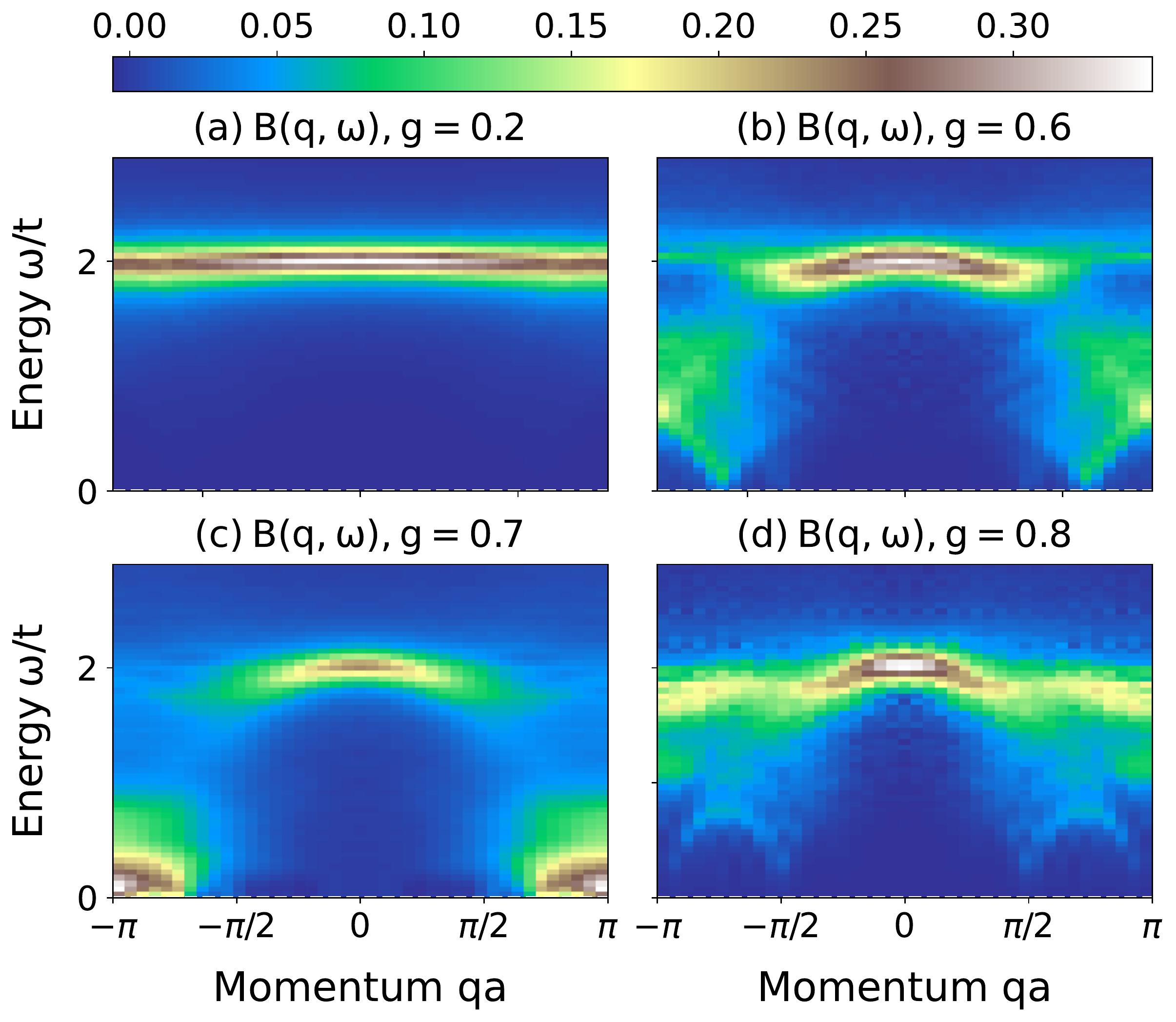}
    \end{minipage}\hfill
    \begin{minipage}{0.49\textwidth}
    \centering
    \includegraphics[width=\columnwidth]{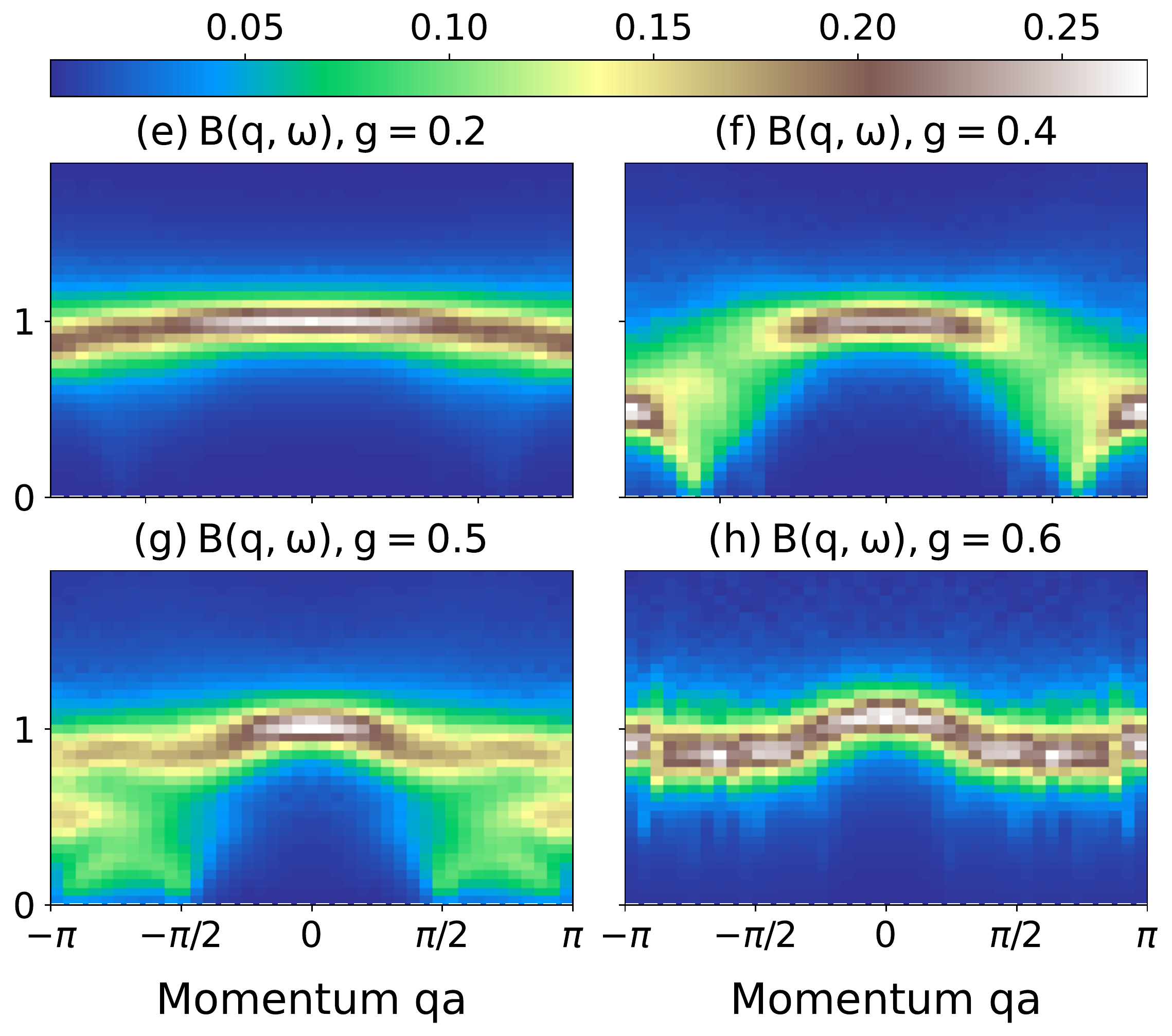}
    \end{minipage}    
    \caption{Phonon spectral function $B(q,\omega)$ for the doped Hubbard-SSH model. The left set of panels shows results for varying SSH coupling $g$ as indicated in each panel, a fixed $U=8t$ and $\Omega=2t$, and an average filling of $n=0.75$ (corresponding to $25\%$ hole doping). The right set of panels shows similar results for fixed $U=8t$, $\Omega=t$, and $n = 0.75$. All spectra were computed using $L =40$ site chains. Here, for $U=8t, \Omega=2t$ and $L=24$ the critical coupling is at $g_\mathrm{c}\approx0.709$ and for $U=8t, \Omega=t$, $g_\mathrm{c}\approx 0.494$.}
    \label{fig:phonon_SF_w2_w1}
\end{figure*}

Figures~\ref{fig:Akw_U8_w2_w1}(e)-(h) show similar results for $U=8t$ and $\Omega=t$. The critical coupling is reduced to $g_\mathrm{c} \approx 0.494$ for these parameters due to the softer harmonic lattice potentials. Nevertheless, the same behavior with increasing coupling is observed but with additional band renormalizations appearing at higher multiples of the phonon energy. For example, Fig.~\ref{fig:Akw_U8_w2_w1}(f) shows results for $g=0.4$, where the characteristic triangular structure of spinon and holon bands has been broken up by the band renormalizations appearing at $\Omega$ and $2\Omega$. For $g=0.5~(\gtrapprox g_\mathrm{c})$ [Fig.~\ref{fig:Akw_U8_w2_w1}(g)], the spectrum already begins to resemble the ones for the static dimerized case. 

\subsection{Phonon spectral functions}

Next, we present results for the phonon spectral function $B(q,\omega)$. Figure \ref{fig:phonon_SF_w2_w1} shows the results for the same parameters used in the previous section following the layout of Fig.~\ref{fig:Akw_U8_w2_w1}. For weak coupling, $g=0.2$ [Figs. \ref{fig:phonon_SF_w2_w1}(a) \& (e)], the spectra consist of a single weakly dispersing peak centered near $\omega=\Omega=2t$ and $t$,  respectively, as expected for an optical phonon branch. However, both curves also exhibit a weak softening near the Brillouin zone boundary, which is more apparent in the $\Omega=t$ spectra. Such softening effects are similar to what is observed for the half-filled Holstein model, where charge-density-wave correlations develop near $q = \pi$~\cite{Weber2015phonon}. As $g$ increases, the phonon dispersion softens more significantly, leading to soft zero-energy modes at $q=\pm 2k_\mathrm{F}$ [Figs.~\ref{fig:phonon_SF_w2_w1}(b) \& \ref{fig:phonon_SF_w2_w1}(f)]. We also observe a spectral weight depletion at the bare phonon energy, which seems inconsistent with an avoided level crossing picture between the flat phonon branch and particle-hole charge excitations often invoked to understand the dispersion softening in the Holstein model~\cite{Weber2015phonon}. As the coupling approaches the critical value $g_\mathrm{c}$ [Fig.~\ref{fig:phonon_SF_w2_w1}(c)], more spectral weight is transferred to low energies at the zone boundaries, with additional weight concentrating at low energy near $q=\pm \pi$. For $g>g_\mathrm{c}$ [Fig.~\ref{fig:phonon_SF_w2_w1}(d), (g), \& (h)],  spectral weight is transferred back to high energies, and the low-energy gapless excitation disappears. This behavior likely reflects the formation of a new doubled unit cell once the dimerized state forms.  

\begin{figure*}[t]
    \begin{minipage}{0.49\textwidth}
    \centering
    \includegraphics[width=\columnwidth]{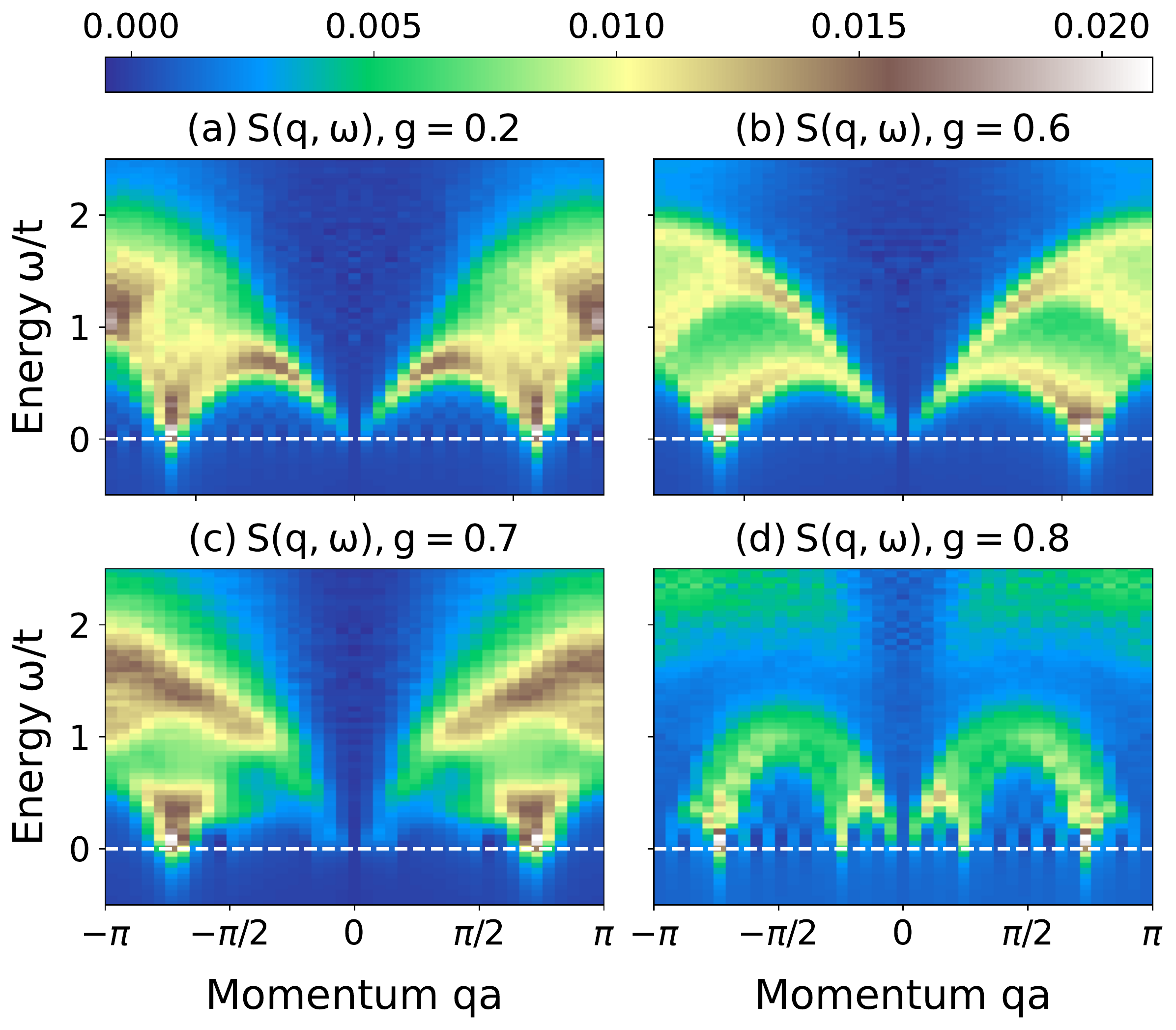}
    \end{minipage}\hfill
    \begin{minipage}{0.49\textwidth}
    \centering
    \includegraphics[width=\columnwidth]{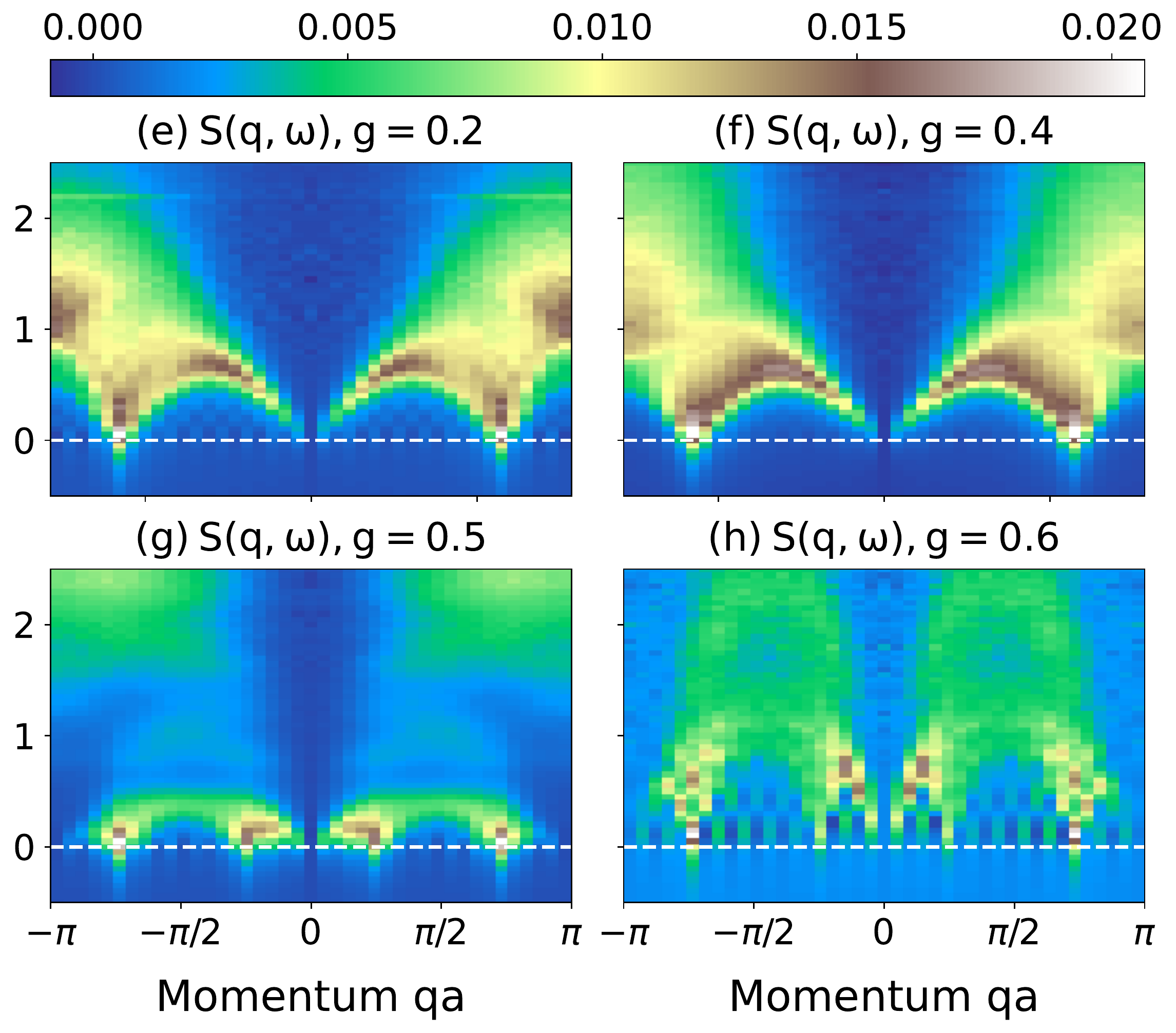}
    \end{minipage}    
    \caption{Dynamical spin structure factor $S(q,\omega)$ for the doped Hubbard-SSH model. The left set of panels shows results for varying SSH coupling $g$ as indicated in each panel, a fixed $U=8t$ and $\Omega=2t$, and an average filling of $n=0.75$ (corresponding to $25\%$ hole doping). The right set of panels shows similar results for fixed $U=8t$, $\Omega=t$, and $n = 0.75$. All spectra were computed using $L =40$ site chains. Here, for $U=8t, \Omega=2t$ and $L=24$ the critical coupling is at $g_\mathrm{c}\approx0.709$ and for $U=8t, \Omega=t$, $g_\mathrm{c}\approx0.494$.}
    \label{fig:Skw_U8_w2_w1}
\end{figure*}

Interestingly, our results in Figs.~\ref{fig:phonon_SF_w2_w1}(b) and (f) bear a strong resemblance to the anomalous softening of the bond-stretching modes observed in the high-T$_\mathrm{c}$ cuprates near the CDW ordering vector~\cite{ChaixNatPhys2017, Miao2019formation}. Here, however, we observe a strong asymmetry of the spectral intensity for $q>q_{CDW}$ as opposed to $q<q_\mathrm{CDW}$ with a maximum intensity at about $\omega\simeq\Omega/2$ for $\Omega=t$ [see Fig.~\ref{fig:phonon_SF_w2_w1}(f)]. This difference may be related to the fact that the RIXS intensity is weighted by the bare $e$-ph coupling constant~\cite{Ament2011determining}, which further modulates the intensity of the phonon features.

\subsection{Dynamical spin structure factors}

We now examine the effects of the SSH interaction on the magnetic excitations encoded in the dynamical spin structure factor $S(q,\omega)$. 
Figure~\ref{fig:Skw_U8_w2_w1} shows DMRG results for $S(q,\omega)$ for an $L=40$ site chain with $\braket{n}=0.75$. Results are shown here for the same parameters used in Fig.~\ref{fig:Akw_U8_w2_w1}, and the panels follow in a one-to-one correspondence with the previous figures. As with the single-particle response functions, $S(q,\omega)$ is weakly modified for small $e$-ph couplings [$g=0.2$, Figs.~\ref{fig:Skw_U8_w2_w1}(a) \& (e)] and closely resemble the spinon continuum typical for a 1D doped Hubbard chain in the Heisenberg limit~\cite{PhysRevB.55.12510, NoceraPRB2016}. Specifically, spectral weight is focused in a sharp peak at the lower boundary of the continuum while the spectrum is gapless at $q=\pm\: 2k_\mathrm{F}$ \cite{PhysRevB.96.195106}. 

As the $g$ increases [Fig.~\ref{fig:Skw_U8_w2_w1}(b)], the lower-energy spin excitations begin to soften, and the spectral weight of the continuum is spread out over a larger range of energy. As $g$ approaches the critical coupling $g_\mathrm{c}$
[Fig.~\ref{fig:Skw_U8_w2_w1}(c)] the spinon continuum near the lower boundary appears to break up, with weight transferred to features at higher energy close to the boundary of the original spinon continuum. 
We find no evidence for the opening of a spin gap for $g<g_\mathrm{c}$, which suggests that the SSH coupling does not drive the system to a spin-gapped Luther-Emery liquid state, expected for the doped Hubbard model at \emph{negative} $U$~\cite{giamarchi2004quantum}, where on-site pairing is expected to dominate. Finally, a more significant reorganization of the magnetic excitations occurs once $g > g_\mathrm{c}$  [Fig.~\ref{fig:Skw_U8_w2_w1}(d)]. For example, the two-spinon continuum is no longer apparent for $g=0.8>g_\mathrm{c}$ [Fig.~\ref{fig:Skw_U8_w2_w1}(d)], which is well within the dimerized regime. Instead, the low-energy magnetic excitations form a relatively sharp inverted parabolic structure that crosses $\omega = 0$ at $q \approx \pm 2k_\mathrm{F}$ and $\pm 3/8 \:(2k_\mathrm{F})$. At the same time, the high-energy weight becomes incoherent and is pushed to energies well above the boundaries of the original spinon continuum. These spectra resemble the magnetic excitation spectrum obtained for the Hubbard dimer model in Fig.~\ref{fig:limits}. The reorganization of the magnetic excitations thus reflects the transition from a doped 1D Hubbard chain to a chain of connected Hubbard dimers. 

Similar results follow for $U=8t, \Omega=t$, for $g \ge 0.5$, as shown in Figs.~\ref{fig:Skw_U8_w2_w1}(g)-(h). 

We end this section by commenting on the changes in the spin velocity $v_{\sigma}$ induced by the SSH coupling, which can be estimated from the slope of the excitation energies in the $S(q,\omega)$ as $q, \omega\rightarrow 0$. Here, we restrict ourselves to the region $g<g_\mathrm{c}$. In the anti-adiabatic regime $\Omega=2t$, we find that the spin velocity progressively increases as a function of the $e$-ph coupling ($v_{\sigma}\approx 0.76$ for $g=0.2$ and $g=0.4$, while $v_{\sigma}\approx 0.82$ for $g=0.6$, $v_{\sigma}\approx 0.98$ for $g=0.7$). This result suggests that weak SSH coupling increases the spinon bandwidth, at least in the anti-adiabatic regime $\Omega\gg t$. For $\Omega=t$, we instead observe that the spin velocity drops quickly by increasing the SSH \emph{e}-ph coupling strength ($v_{\sigma}\approx 0.76$ for $g=0.2$ and $v_{\sigma}\approx 0.63$ for $g=0.4$). This second observation suggests that there are significant retardation effects entering the magnetic properties of the single-orbital Hubbard-SSH model for realistic values of the phonon energy $\Omega\lesssim t$, which may be necessary for understanding the interplay between $e$-ph coupling and magnetism in these materials. In this context, we note that a recent determinant quantum Monte Carlo study of a multi-orbital corner-shared CuO$_4$ chain model~\cite{LiPreprint2022} found that SSH coupling to high-energy phonons in the anti-adiabatic limit suppresses the superexchange coupling for $g < g_\mathrm{c}$. Collectively, these results suggest that the way in which the SSH interaction modifies the magnetic properties of oxides can depend strongly on the phonon energy and whether one adopts a single- or multi-band description of the system. 

\begin{figure*}[t]
    \begin{minipage}{0.49\textwidth}
    \centering
    \includegraphics[width=\columnwidth]{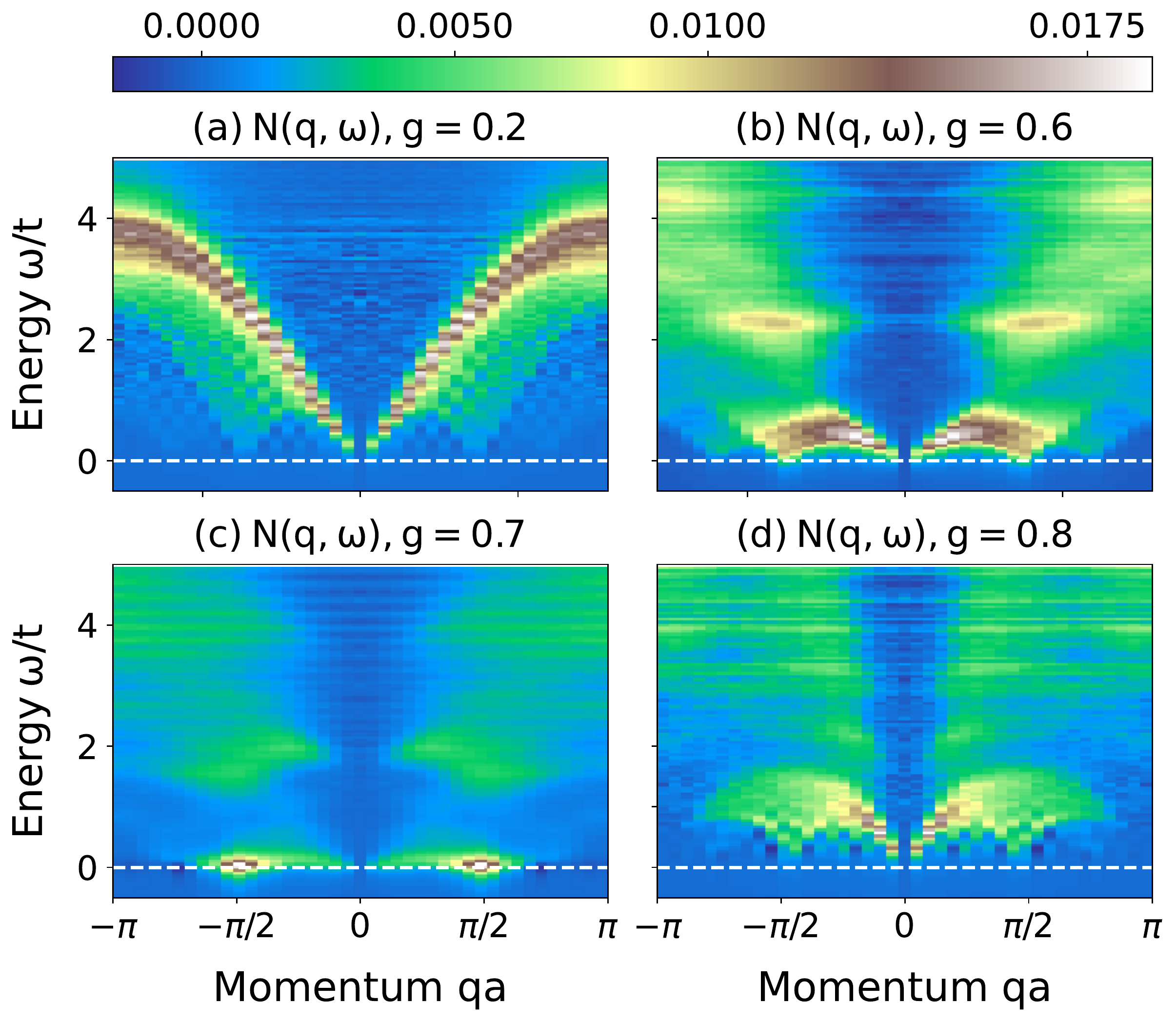}
    \end{minipage}\hfill
    \begin{minipage}{0.49\textwidth}
    \centering
    \includegraphics[width=\columnwidth]{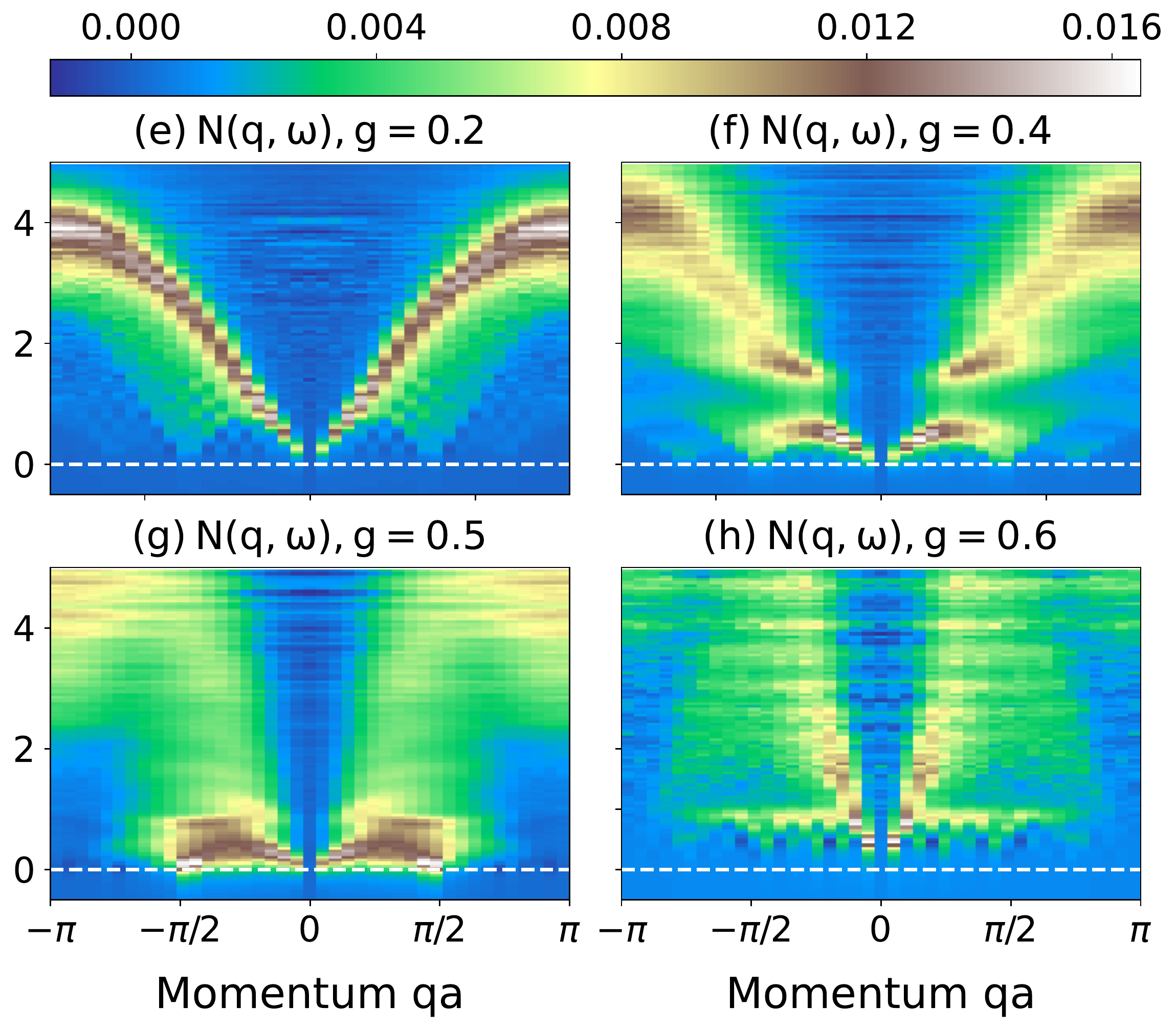}
    \end{minipage}    
    \caption{Dynamical charge structure factor $N(q,\omega)$ for the doped Hubbard-SSH model. The left set of panels shows results for varying SSH coupling $g$ as indicated in each panel, a fixed $U=8t$ and $\Omega=2t$, and an average filling of $n=0.75$ (corresponding to $25\%$ hole doping). The right set of panels shows similar results for fixed $U=8t$, $\Omega=t$, and $n = 0.75$. All spectra were computed using $L =40$ site chains. Here, for $U=8t, \Omega=2t$ and $L=24$ the critical coupling is at $g_\mathrm{c}\approx0.709$ and for $U=8t, \Omega=t$, $g_\mathrm{c}\approx0.494$.}
    \label{fig:Nkw_U8_w2_w1}
\end{figure*}

\subsection{Dynamical charge structure factors}
Finally, we present results for the dynamical charge structure factors. Fig.~\ref{fig:Nkw_U8_w2_w1} shows $N(q,\omega)$ for the same parameters used in the earlier figures, following the same panel labeling. As with the spectral function and dynamical spin structure factor, the dynamical charge structure factor is only weakly dressed for weak $e$-ph coupling $g=0.2$ for $\Omega=2t$ [Fig.~\ref{fig:Nkw_U8_w2_w1}(a)] and $\Omega = t$ [Fig.~\ref{fig:Nkw_U8_w2_w1}(e)]. For example, for $\Omega = 2t$ and $g = 0.2$,  the main peak in $N(q,\omega)$ has a kink-like structure and broadens as it crosses the phonon energy. (This same renormalization is less prominent in the $\Omega = t$ data because the phonon energy is closer to the lower boundaries of the charge excitation continuum.) These renormalizations become more pronounced as $g$ increases toward $g_\mathrm{c}$, with additional features forming at multiples of the phonon energy. This behavior results from creating the Franck-Condon shake-off states in the spectral functions shown in Fig.~\ref{fig:Akw_U8_w2_w1}. We envision that such spectral features could be experimentally observed in future experiments on doped cuprate chains at the oxygen $K$-edge~\cite{Schlappa2018probing}, where charge excitations are expected to dominate at low energy.

As the coupling increases beyond the critical coupling $g>g_\mathrm{c}$, the system transitions into the dimerized state, and the low-energy $N(q,\omega)$ spectra more closely resemble the static dimerized limit shown in Fig.~\ref{fig:limits}(f). However, we also observed a significant amount of incoherent spectral weight at higher energies, which is absent from the static calculations. This difference highlights that the dimerization process involves a substantial coupling between the lattice and the electrons. In this case, the electrons in the dimerized structure should be viewed as (bi)polarons, where carriers are bound to the sites that form the short bond with a cloud of bond phonons~\cite{NoceraPRB2022}. 

Finally, we end this section commenting on the charge velocity $v_\rho$, which can be extracted from the slope of the excitation energies in $N(q,\omega)$ as $q,\omega\rightarrow 0$. As with the spin excitations, we restrict ourselves to $g<g_\mathrm{c}$. Contrary to the case of spin excitations, we observe that both for $\Omega=2t$ and $\Omega=t$, the charge velocity progressively reduces by increasing the \emph{e}-ph coupling strength (for $\Omega=2t$, $v_{\rho}\approx1.72,1.4,1.145,0.76$ for $g=0.2,0.4,0.5,0.6$ while for $\Omega=t$, $v_{\rho}\approx1.68,0.76$ for $g=0.2,0.4$), pointing towards a strong holon bandwidth renormalization induced by the SSH \emph{e}-ph coupling. A similar reduction occurs in the doped 1D Hubbard-Holstein model~\cite{NoceraPRB2014}.


\section{Discussion and Conclusions}
We have studied the doped one-dimensional SSH-Hubbard model using the density matrix renormalization group method and presented results for its ground state correlations, single-particle electron and phonon spectral functions, and its two-particle dynamical spin and charge structure factors. 

The SSH interaction modulates the nearest neighbor hopping integrals at linear order in the displacements. Due to this linear approximation, the interaction can dimerize the effective nearest-neighbor hopping integrals leading  $-t_\mathrm{eff}=-t(1 - A)$ and $-t(1+A)$ alternating along the chain, where $A \approx g\langle X_i-X_{i+1}\rangle$ in a mean-field picture. Importantly, if $A \ge t$, the effective hopping integral along the long bond will have an inverted sign relative to the undistorted lattice. Our results demonstrate that this dimerization persists in a numerically exact treatment of the problem and is accompanied by a significant reorganization of the ground and excited state correlations. 
Because of this dimerization, the spectral properties of the SSH model are substantially different from the more widely studied Holstein model of $e$-ph  coupling in the strong coupling limit. For example, in addition to the expected reduction of spinon (observed for $\Omega=t$ but not for $\Omega=2t$ for $g$ below the critical \emph{e}-ph coupling $g_\mathrm{c}$) and holon bandwidths, the SSH interaction introduces 
spectral features in $A(k,\omega)$ that has no counterpart in the Holstein model. 

It is also important to consider our results in the context of the recent ARPES experiments on the doped 1D spin-chain Ba$_{2-x}$Sr$_x$CuO$_{3+\delta}$~\cite{ChenScience2021}. Excess spectral weight was observed in the back-folded holon bands that could not be accounted for using the standard single-band Hubbard model with local repulsive interaction. Instead, it was found that this additional weight could be recovered if a substantial next-nearest-neighbor  \textit{attractive} interaction $V \sim -t$ was included in the model. Subsequent theoretical works ~\cite{WangPRL2021, wang2022spectral, tang2022traces} have argued that an extended Holstein coupling could account for this additional interaction. However, there is also strong evidence for a connection between the Cu-O bond-stretching phonons and charge order in 2D cuprates ~\cite{Lanzara2001, JohnstonPRB2010, ChaixNatPhys2017}. Therefore, it is natural to wonder whether the corresponding SSH coupling could also be relevant for quasi-1D spin chain cuprates. While our numerical results show a spectral feature [see Fig.~\ref{fig:Akw_U8_w2_w1}(b) and (f)] consistent with the so-called holon folding mode at intermediate SSH coupling, this feature does not have enough spectral intensity to account for the observed weight in Ba$_{2-x}$Sr$_x$CuO$_{3+\delta}$~\cite{ChenScience2021}. 

Our results for the phonon spectral function also highlight some interesting differences between the SSH model's dimerization process and more conventional nesting-driven Peierls scenarios in 1D. In the latter case, the dimerization process is driven by perfect Fermi surface nesting, and one would expect a sharp Kohn anomaly to develop in the phonon dispersion, where the modes near $q\sim 2k_\mathrm{F}$ soften to zero. While our results exhibit a softening at this wave vector, we also observe a significant softening of the modes near the zone boundary. Interestingly, further increases in the coupling cause the phonon spectrum to harden, thus eliminating the soft mode once the dimerized state has formed. 

Finally, we summarize our main results from the spin and charge dynamical structure factors. The spin excitations of the hole-doped Hubbard-SSH chain for $g < g_\mathrm{c}$ show at low energy the main features of the spectrum of a doped Hubbard chain with gapless excitations at $q=\pm 2k_{\mathrm{F}}$ with minor spectral weight reorganizations at higher energies. For larger \emph{e}-ph couplings corresponding to the sign inversion of the effective hopping, they interestingly display a strong depletion of spectral weight at intermediate energies (of the order of the phonon energies) while the remaining spectral weight is pushed at lower and higher energies, respectively.
In the charge dynamical structure factors, and for $g$ below $g_\mathrm{c}$, a strong depletion of spectral weight at intermediate to high energies (of the order of the phonon energies, and also multiple of it) appears at a moderate \emph{e}-ph coupling strength. This characteristic signature of the SSH coupling could be experimentally verified in future RIXS experiments on doped cuprate chains at oxygen $K$-edge~\cite{Schlappa2018probing}, where non-spin-flip charge excitations are expected to dominate the signal. For $g>g_\mathrm{c}$, instead, the spectral weight becomes largely incoherent at high energy while showing the same spectral features of a statically dimerized model at low energies. 

\section*{Acknowledgments}
We thank G. G. Batrouni, M. Berciu, and A. Feiguin for useful discussions. This work was supported by the National Science Foundation under Grant No. DMR-1842056. A.~N. acknowledges support from the Max Planck-UBC-UTokyo Center for Quantum Materials and Canada First Research Excellence Fund (CFREF) Quantum Materials and Future Technologies Program of the Stewart Blusson Quantum Matter Institute (SBQMI), and the Natural Sciences and Engineering Research Council of Canada (NSERC). 

\bibliography{references}
\end{document}